\documentclass[review]{elsarticle}

\usepackage[hidelinks]{hyperref}

\journal{}

\usepackage{amsmath,amsfonts}
\usepackage{algorithm}
\usepackage{algorithmic}
\usepackage{setspace}
\usepackage{array}
\usepackage[caption=false]{subfig}
\usepackage{textcomp}
\usepackage{stfloats}
\usepackage{url}
\usepackage{verbatim}
\usepackage{graphicx}
\usepackage{balance}

\usepackage[margin=1in]{geometry}

\usepackage{amsmath,amssymb}
\usepackage{graphics}
\usepackage[caption=false]{subfig}

\usepackage{pgfplots}

\definecolor{mypurple}{rgb}{0.5961,0.3059,0.6392}%
\definecolor{mygreen}{rgb}{0.7020,0.8706,0.4118}%
\definecolor{myorange}{rgb}{0.9843,0.5020,0.4471}%
\definecolor{myblue}{rgb}{0.00000,0.44700,0.74100}%

\usepackage{nicefrac}
\usepackage{xcolor}
\usepackage[caption=false]{subfig}
\usepackage[capitalize]{cleveref}

\usepackage{tikz}
\usepackage{pgfplots}

\definecolor{mypurple}{rgb}{0.5961,0.3059,0.6392}\definecolor{mygreen}{rgb}{0.7020,0.8706,0.4118}\definecolor{myorange}{rgb}{0.9843,0.5020,0.4471}\definecolor{myblue}{rgb}{0.00000,0.44700,0.74100}

\newcommand{\myfigsize}{0.7}
\tikzset{every plot/.style=thick}
\pgfplotsset{
    is legend/.append style={
font=\footnotesize
    }
}

\usepackage[normalem]{ulem}

\definecolor{brightpink}{rgb}{1.0, 0.0, 0.5}

\definecolor{darkgreen}{rgb}{0.01, 0.75, 0.24}

\newcommand{\revise}[1]{{\color{black} #1}}

\newtheorem{theorem}{Theorem}

\newtheorem{assumption}{Assumption}
\newtheorem{remark}{Remark}
\newtheorem{problem}{Problem}

\DeclareMathOperator{\conv}{conv}
\DeclareMathOperator{\argmax}{argmax}

\DeclareMathOperator{\rank}{rank}

\DeclareMathOperator{\nnz}{nnz}

\begin{document}

\begin{frontmatter}

\title{Smoothed Separable Nonnegative Matrix Factorization}

\author[mons]{Nicolas Nadisic\corref{cor}}
\cortext[cor]{Corresponding author}
\ead{nicolas.nadisic@umons.ac.be}

\author[mons]{Nicolas Gillis}
\ead{nicolas.gillis@umons.ac.be}

\author[paris]{Christophe Kervazo}
\ead{christophe.kervazo@telecom-paris.fr}

\address[mons]{Department of Mathematics and Operational Research\\
Faculté Polytechnique, Université de Mons\\
Rue de Houdain 9, 7000 Mons, Belgium.
}
\address[paris]{LTCI, Télécom Paris, Institut Polytechnique de Paris\\
19 Place Marguerite Perey, 91120 Palaiseau, France.
}

\begin{abstract}
Given a set of data points belonging to the convex hull of a set of vertices, a key problem in linear algebra, signal processing, data analysis and machine learning is to estimate these vertices in the presence of noise. 
Many algorithms have been developed under the assumption that there is at least one nearby data point to each vertex; two of the most widely used ones are vertex component analysis (VCA) and the successive projection algorithm (SPA). 
This assumption is known as the pure-pixel assumption in blind hyperspectral unmixing, and as the separability assumption in nonnegative matrix factorization. 
More recently, Bhattacharyya and Kannan 
(ACM-SIAM Symposium on Discrete Algorithms, 2020) 
proposed an  algorithm for learning a latent simplex (ALLS) that relies on the assumption that there is more than one nearby data point to each vertex.
In that scenario, ALLS is probalistically more robust to noise than algorithms based on the separability assumption.  
In this paper, inspired by ALLS, we propose smoothed VCA (SVCA) and smoothed SPA (SSPA) that generalize VCA and SPA by assuming the presence of several nearby data points to each vertex. 
We illustrate the effectiveness of SVCA and SSPA over VCA, SPA and ALLS on synthetic data sets, on the unmixing of hyperspectral images\revise{, and on feature extraction on facial images data sets}.
In addition, our study highlights new theoretical results for VCA.
\end{abstract}

\begin{keyword}
nonnegative matrix factorization \sep
separability \sep
blind hyperspectral unmixing \sep
pure-pixel search algorithms \sep
latent simplex \sep
simplex-structured matrix factorization 
\MSC[2020] 65F55 \sep 15A23 \sep 68U10
\end{keyword}

\end{frontmatter}


\section{Introduction} \label{sec:intro}

{Low-rank matrix factorization is a widely used class of unsupervised data analysis models. 
Examples include
principal component analysis~\cite{jolliffepca}, 
independent component analysis~\cite{tharwat2020independent}, 
sparse component analysis~\cite{gribonval2010sparse}, 
and nonnegative matrix factorization~\cite{gillis2020book}, 
to cite a few. 
In this paper, 
we focus on the following matrix factorization model:
given a set of data points within the convex hull of a set of vertices,  estimate these vertices in the presence of noise.
}  
This problem can be formulated as follows.
\begin{problem}\label{prob:main}
Given ${X} = WH + N \in \mathbb{R}^{m \times n}$
where $H \in \mathbb{R}^{r \times n}_+$ is column stochastic and $N$ is the noise,
estimate $W \in \mathbb{R}^{m \times r}$.
\end{problem}
\noindent 
Note that $W$, $H$ and $N$ are unknown, only $X$ is given. 
In this paper, we focus on the estimation of $W$. Once $W$ is estimated, the matrix $H$ can be estimated by solving a convex nonnegative least squares problem.
Problem~\ref{prob:main} is sometimes referred to as simplex-structured matrix factorization (SSMF), and generalizes nonnegative matrix factorization (NMF);
see~\cite{abdolali2020simplex} and the references therein. \revise{It is also closely related to semi-NMF that only requires $H$ to be nonnegative but not column stochastic~\cite{ding2008convex}.}  
{Problem~\ref{prob:main} is closely related to the column subset selection problem (see, e.g.,~\cite{derezinski2020improved, shitov2021column} and the references therein), but the additional assumptions allow us to tackle it using convex geometry concepts.} 

In Problem~\ref{prob:main}, the columns of $W$ are the vertices, while the columns of $X$ are noisy data points within the convex hull of the columns of $W$,
\[
\conv(W) = \{ x \ | \ x = Wy, y \geq 0, e^\top y = 1 \},
\]
where $e$ is the vector of all ones of appropriate dimension, {and $^\top$ is the transpose of a vector or a matrix}. In fact, for all $j$, 
\[
{X}(:,j) = WH(:,j) + N(:,j) ,
\] 
 {where $X(:,j)$ denotes the $j$th column of $X$}, 
 $H(:,j) \geq 0$ and $e^\top H(:,j) = 1$ (since $H$ is column stochastic), which means that 
 ${X}(:,j) - N(:,j) \in \conv(W)$ for all $j$. 
To be able to estimate $W$ in Problem~\ref{prob:main}, appropriate assumptions on $W$, $H$ and $N$ are required.
In particular, the following assumptions are necessary~\cite{abdolali2020simplex}: 
\begin{itemize}

\item No column of $W$ is contained in the convex hull of the other columns of $W$, otherwise it is not possible to distinguish it from a data point.

\item The data points must be sufficiently spread within $\conv(W)$, having data points on each facet of $\conv(W)$. This implies some degree of sparsity for~$H$.

\item The noise $N$ must be bounded.

\end{itemize}

To obtain provable or practical algorithms, the above assumptions must be carefully and rigorously complemented. 
Different assumptions on $W$, $H$, and $N$ lead to different models for which different algorithms can be designed; see \cref{sec:ssmf} for a literature review.
We discuss below some applications of such algorithms.

\paragraph*{Applications in data analysis and machine learning}
The papers \cite{bhattacharyya2020finding} and \cite{bakshilearning} describe in details  various applications of solving Problem~\ref{prob:main}; in particular topic modeling via
latent Dirichlet allocation, adversarial clustering and community detection via the mixed membership stochastic block model.
Moreover, Problem~\ref{prob:main} generalizes NMF and as such could be useful for all applications of NMF, such as 
feature extraction in sets of images~\cite{lee1999learning}, 
audio source separation~\cite{ozerov2009multichannel}, chemometrics~\cite{thiel2021comparison}, 
or blind hyperspectral unmixing (HU)~\cite{pauca2006nonnegative, gillis2013sparse}.
Although the model considered in this work is very generic, we are motivated mainly by blind HU, a key problem in remote sensing.
Let us briefly describe this problem;
see the survey papers by~\cite{bioucas2012hyperspectral, ma2014signal} and the references therein for more details.

\paragraph*{Blind hyperspectral unmixing}

A hyperspectral image (HSI) is a picture of a scene acquired within a large number of spectral bands (usually between 100 and 200). Thus, for each pixel a precise electromagnetic spectrum is recorded, which gives an information concerning the materials present in the pixel; specifically, about their reflectances (fraction of incoming light they reflect) and/or their emissivity (which is due to the fact that the materials usually have non-zero temperatures). Unfortunately, despite their high spectral resolution, hyperspectral sensors generally have a low spatial resolution; as such, the spectrum recorded for each pixel might not correspond to the one of a single material, but rather to a mixture of the spectra of the different materials present within the pixel.

Given an HSI, blind HU thus aims to recover the set of materials present in the image, called endmembers, along with the abundances of each endmember in each pixel. The standard model used to solve blind HU is the linear mixing model. It assumes that the spectral signature of each pixel is a linear combination of the spectral signatures of the endmembers, where the weights of the linear combination are the abundances of the endmembers in the pixel. 
Typically, we represent a hyperspectal image as a matrix $X$, where the $j$th column, $X(:,j)$, corresponds to the spectral signature of the $j$th pixel in the scene.
If the spectral signatures of the endmembers are also collected as the columns of a matrix $W$, then according to the linear mixing model, the $j$th pixel can be written as $X(:,j)  \approx
\sum_{k=1}^r W(:,k) H(k,j) + N(:,j)$, where $H(k,j)$ is the abundance of the $k$th endmember in the $j$th pixel, and $N(:,j)$ represents the noise and model misfit.
This is exactly the setup of Problem~\ref{prob:main}.

\paragraph*{Pure-pixel assumption} 
An important class of algorithms to solve blind HU are \emph{pure-pixel search} algorithms.
They rely on the assumption that for each endmember, there is at least one pixel in which this endmember appears almost alone, that is, purely, so that the endmember signature is close to the one of the corresponding pure pixel. 
Two of the most effective and widely used pure-pixel search algorithms are vertex component analysis (VCA)~\citep{nascimento2005vertex} and the successive projection algorithm (SPA)~\citep{Araujo01} which will be described in Section~\ref{sec:ssmf}.

\paragraph*{Terminology}  
In the NMF literature, the pure-pixel assumption is referred to as separability, and the corresponding algorithms are referred to as separable NMF algorithms, or near-separable NMF algorithms. 
Geometrically, the pure-pixel assumption requires that there is a data point (that is, a column of $X$), close to each vertex  (that is, to each column of $W$).
We will refer to such a data point as pure when the corresponding column of $H$ is a unitary vector (that is, a column of the identity matrix).

\paragraph*{Contribution and outline}

When the separability assumption holds, there are typically more than one data point close to each vertex. In this paper, we leverage this observation by adapting VCA and SPA, providing two new algorithms, namely smoothed VCA (SVCA) and smoothed SPA (SSPA). 
The idea is to aggregate (for instance average) several data points around a vertex to obtain a better estimate of that vertex.
In practice, it enables our two smoothed algorithms to achieve much better separation results than their non-smoothed counterparts, and to be more tolerant to noise.   

SVCA can also be interpreted as an adaptation of the  algorithm for learning a latent simplex (ALLS) proposed by~\cite{bhattacharyya2020finding}, with two major modifications {that we detail in \cref{sec:svca}}. 
This observation allows us to provide theoretical guarantees for SVCA, and hence VCA which is a special case of SVCA. To the best of our knowledge, this is the first time a theoretical guarantee for VCA is provided in the presence of noise.

We believe that this work paves the way for a whole new branch of smoothed separable NMF algorithms. 
Although we focus in this work on VCA and SPA, two of the most well-known separable NMF algorithms, it is straightforward to extend the methodology to more recent algorithms such as SNPA~\citep{gillis2014successive}.

The paper is organized as follows.
In \cref{sec:ssmf},
we summarize the literature for solving Problem~\ref{prob:main}, with a focus on three algorithms, namely VCA, SPA and ALLS.
In \cref{sec:svca}, we propose SVCA which is equivalent to applying VCA on a smoothed data set. 
SVCA is similar to ALLS, but two key differences make it empirically more efficient than ALLS. 
In \cref{sec:sspa}, we propose SSPA which adapts SPA in the presence of multiple pure pixels.
In \cref{sec:nex}, we show on synthetic and real-world hyperspectral data sets that SVCA and SSPA outperform VCA, SPA and ALLS in the presence of multiple pure pixels.

{
\paragraph{Notation} Given an $m$-by-$n$ real matrix $X \in \mathbb{R}^{m \times n}$, 
$X^\top$ is its transpose, 
$X(:,j)$ is its $j$th column, 
$X(i,:)$ is its $i$th row, 
$X(i,j)$ is its entry at position $(i,j)$, 
\mbox{$X(:,\mathcal{K})$} is the submatrix made of the columns of $X$ in the index set $\mathcal{K}$, and similarly for \mbox{$X(\mathcal{K},:)$} for the rows. 
The $i$th singular value of $X$ is denoted $\sigma_i(X)$, while 
$K(W)  =  \max_j \| W(:,j) \|_2$. 
If a matrix $X$ is component-wise nonnegative, we write $X \geq 0$.
The vector of all ones is denoted $e$.
The $m$-by-$m$ identity matrix is denoted $I_m$.
}

\section{Simplex-structured matrix factorization} \label{sec:ssmf}

In this section, we describe existing models and algorithms to solve Problem~\ref{prob:main}, that is, to solve SSMF, in order to identify the vertices of the convex hull of a set of data points, ${X}$.
We focus on two key models, upon which our contribution is built:

\begin{enumerate}

\item Separable NMF: it assumes there is one data point close to each vertex of $\conv(W)$,
and that the noise added to each data point is bounded;
see \cref{sec:model1}. 
Some authors refer to this model as near-separable NMF.

\item  Learning a latent simplex: it is motivated by machine learning applications. It assumes that there is more than one data point close to each vertex of $\conv(W)$, but it allows much larger noise levels as it only requires
the $\ell_2$ norm of $N$ to be bounded, instead of each individual column; see \cref{sec:model2}.

\end{enumerate}

Other models and algorithms exist to tackle Problem~\ref{prob:main} relying on different assumptions.
It is worth mentioning minimum-volume NMF~\citep{fu2015blind}, where $W$ is regularized such that its convex hull $\conv(W)$ has the smallest possible volume, 
facet-based identification algorithms that identify the facets of $\conv(W)$ from which its vertices are recovered~\citep{GeZouNMF,  lin2015fast, lin2017maximum, abdolali2020simplex}, 
and probabilistic simplex component analysis~\citep{wu2021probabilistic} that relies on a probabilistic model on the data (the columns of $H$ are sampled using the Dirichlet distribution, and the entries of $N$ using i.i.d.\ Gaussian noise). 
Discussing these approaches in detail is out of the scope of this article, whose focus is on the separable NMF problem.

\subsection{Model 1: Separable NMF} \label{sec:model1}

As already mentioned, an important class of NMF algorithms relies on the separability assumption, defined as follows.
\begin{assumption}[separability]  \label{ass:sep}
In Problem~\ref{prob:main}, there exists an index set $\mathcal{K}$ of cardinality $r$ such that $H(\mathcal{K},:) = I_r$ where $I_r$ is the $r$-by-$r$ identity matrix. \end{assumption}
Under this assumption, solving Problem~\ref{prob:main} amounts to recover $\mathcal{K}$ such that
\[
{X}(:,\mathcal{K})
\; = \;
W + N(:,\mathcal{K}) 
\quad \approx  \quad W. 
\]
{In blind HU, separability is known as the pure-pixel assumption.}

The early algorithms building on this assumption emerged in the blind HU community. They include {pixel purity index} (PPI)~\cite{boardman1995mapping}, N-FINDR~\cite{winter1999n}, and vertex component analysis (VCA)~\cite{nascimento2005vertex}.
Most of these algorithms were developed based on convex geometry concepts.
These early works however did not analyze noise robustness, and in fact they are not guaranteed to recover the endmembers in the presence of noise.

In analytical chemistry, Problem~\ref{prob:main} is closely related to the problem of self-modeling curve resolution~\citep{jiang2004principles}. 
As in blind HU, several algorithms were developed based on geometry concepts; in particular the successive projection algorithm (SPA)~\citep{Araujo01}.

More recently, and motivated by applications in machine learning (in particular, topic modeling where pure data points are referred to as anchor words),
\cite{AGKM11} introduced the first provably robust separable NMF algorithms.
Their robustness is deterministic: under some conditions, 
their algorithm is guaranteed to recover an approximation of the vertices.
Arora et al.\ were not aware of the algorithms developed within the blind HU literature.
Many provably robust algorithms have followed this seminal paper, including algorithms that use linear programming~\citep{recht2012factoring, gillis2013robustness, gillis2014robust},
a generalization of SPA~\citep{gillis2013fast},
fast anchor words~\citep{arora2013practical},
and
the successive nonnegative projection algorithm (SNPA)~\citep{gillis2014successive}.
These deterministically robust algorithms guarantee that, in the presence of noise, the vertices are recovered, up to some error bounds that depend on the noise level and the conditioning of $W$;
see \cref{sec:spa} for such a result for SPA.
We refer the interested reader to~\cite[Chapter 7]{gillis2020book} for a detailed discussion and comparison of these algorithms.

In the following, we describe in more detail VCA and SPA that will be instrumental in proposing our new algorithms, smoothed VCA in \cref{sec:svca} and smoothed SPA in \cref{sec:sspa}.

\subsubsection{Vertex component analysis} \label{sec:vca}

VCA~\citep{nascimento2005vertex} is a greedy separable NMF algorithm, that is, it identifies the indices of the subset $\mathcal{K}$ sequentially.
The index set is initialized with $\mathcal{K} = \emptyset$. 
At each of the $r$ iterations of VCA,
a random direction belonging to the subspace spanned by the 
$r$ top left singular vectors of $X$ is generated (this is equivalent to working with the best rank-$r$ approximation of $X$, and hence filters the noise).  
This direction is then projected onto the orthogonal complement of ${X}(:,\mathcal{K})$, and the index of the column of ${X}$ that maximizes the absolute value of the inner product with that direction is added to $\mathcal{K}$. 
Algorithm~\ref{algo:VCA} summarizes VCA.
\algsetup{indent=2em}
\begin{algorithm}[ht!]
\caption{Vertex Component Analysis (VCA) \citep{nascimento2005vertex} \label{algo:VCA}}
\begin{algorithmic}[1]

\REQUIRE The matrix $X \in \mathbb{R}^{m \times n}$,
the number~$r$ of columns to extract.  \vspace{0.1cm}

\ENSURE Index set $\mathcal{K}$ of cardinality $r$ such that $X \approx X(:,\mathcal{K}) H$ for some $H \geq 0$.
    \medskip

\STATE Let $\mathcal{K} = \emptyset$, $P^\bot = I_m$, $V = [\;]$.  \vspace{0.1cm}

\STATE {Let the column of $Y \in \mathbb{R}^{m \times r}$ be the top $r$ left singular vectors of $X$}. \vspace{0.1cm}

\FOR {$k=1$ : $r$}   \vspace{0.1cm}

\STATE Pick a random direction $d_k \in \mathbb{R}^{m}$ in the subspace spanned by $Y$,
e.g.,  $d_k \sim Y \mathcal{N}(0,I_r)$ {where 
$\mathcal{N}(0,I_r)$ is the normal distribution of mean $0$ and covariance matrix $I_r$.} \vspace{0.1cm}

\STATE Compute $u_k = (d_k^\top P^\bot) X \in \mathbb{R}^{n}$, and let
$j_k = \argmax_{1 \leq j \leq n} |u_k(j)|$. \vspace{0.1cm}

\STATE Let $\mathcal{K} = \mathcal{K} \cup \{ j_k \}$. \vspace{0.1cm}

\STATE  Update the projector $P^\bot$ onto the orthogonal complement of \mbox{$W = X(:,\mathcal{K})$}: \label{li:updtproj}
\begin{align*}
& v_k = \frac{P^\bot X(:,j_k)}{\| P^\bot X(:,j_k) \|_2},\\
& V = [V \, v_k],\\
& P^\bot \leftarrow \left( I_m - VV^\top \right).    
\end{align*}

\ENDFOR

\end{algorithmic}
\end{algorithm}

The computational cost of VCA is $\mathcal{O}(r \nnz(X))$ operations, where $\text{nnz}(X)$ is the number of non-zero entries of $X$.
The main cost is to 
compute $Y$ which can be done efficiently using the subspace power iteration, in $\mathcal{O}(\nnz(X)r)$ operations, and to compute the products
$(d_k^\top P^\bot) X$ at each of the $r$ iterations.

An important drawback of VCA is that it is not guaranteed to be deterministically robust to noise. In other words, for any noise $N$ such that a data point goes outside $\conv(W)$, there is a non-zero probability that VCA extracts this point.
The reason is that VCA uses a linear function to identify the vertices of $\conv(W)$; see~\cite[Chapter 7.4]{gillis2020book} for a numerical example.

\subsubsection{Successive projection algorithm} \label{sec:spa}

SPA~\citep{Araujo01} is very similar to VCA. The only difference is in the selection step, when adding an index to $\mathcal{K}$.
SPA selects the column of $P^\bot {X}$ with maximum $\ell_2$ norm; see Algorithm~\ref{algo:SPA}.

\algsetup{indent=2em}
\begin{algorithm}[ht!]
\caption{Successive Projection Algorithm (SPA) \citep{Araujo01} \label{algo:SPA}}
\begin{algorithmic}[1]

\REQUIRE The matrix $X \in \mathbb{R}^{m \times n}$,
the number~$r$ of columns to extract.  \vspace{0.1cm}

\ENSURE Index set $\mathcal{K}$ of cardinality $r$ such that $X \approx X(:,\mathcal{K}) H$ for some $H \geq 0$.
    \medskip

\STATE Let $\mathcal{K} = \emptyset$, $P^\bot = I_m$, $V = [\;]$.  \vspace{0.1cm}

\STATE Let $u_1(j) = \| X(:,j) \|_2^2$ for all $j$.

\FOR {$k=1$ : $r$}   \vspace{0.1cm}

 \STATE Let $j_k = \argmax_{1 \leq j \leq n} u_k(j)$. (Break ties arbitrarily, if necessary.)

\STATE Let $\mathcal{K} = \mathcal{K} \cup \{ j_k  \}$. \vspace{0.1cm}

\STATE {Update the projector $P^\bot$ (as in step \ref{li:updtproj} of Algorithm~\ref{algo:VCA}).}

\STATE  Update the squared norms of the columns of $P^\bot X$:
 for all $j$,
\[
u_{k+1}(j) = u_{k}(j) - v_k^\top X(:,j) = \|P^\bot X(:,j)\|_2^2.
\]

\ENDFOR

\end{algorithmic}
\end{algorithm}

It is interesting to note that VCA is equivalent to SPA if the direction $u_k$ randomly chosen at each step is instead taken as the column of the residual $P^\bot X$ with maximum $\ell_2$ norm. We will use this observation for our proposed algorithm, smoothed SPA.

\begin{remark}[On the projection $P^\bot$] As opposed to VCA, SPA does not work on the subspace spanned by the first $r$ singular vectors of $X$, and hence we stick to this variant in this paper. 
However, in practice, projecting the data onto this subspace allows noise filtering and typically leads to better numerical performance. 
\end{remark}

\paragraph*{Robustness of SPA}
As opposed to VCA, SPA is deterministically robust to noise, provided the following assumption on top of separability (Assumption~\ref{ass:sep}):
\begin{assumption}[column-wise bounded noise]
\label{ass:cwn}
In Problem~\ref{prob:main}, the noise satisfies $\|N(:,j)\|_2 \leq \epsilon$ for all~$j$ for some $\epsilon > 0$ sufficiently small.
\end{assumption}

Let us state the robustness result for SPA.
\begin{theorem}\cite[Theorem 3]{gillis2013fast} \label{th:spa}
Let ${X} = WH+N$ as in Problem~\ref{prob:main}, and let Assumptions~\ref{ass:sep} and~\ref{ass:cwn} be satisfied, that is, $W = X(:,\mathcal{K}^*)$ for some index set $\mathcal{K}^*$ of cardinality $r$, and $\| N(:,j)\|_2 \leq \epsilon$ for all $j$ where $\epsilon \leq \mathcal{O}\left( \frac{\sigma_r^3(W)}{\sqrt{r} K(W)^2} \right)$ {with $K(W)  =  \max_j \| W(:,j) \|_2$}. 
Let also the $r$th singular value of $W$ be positive, that is, $\sigma_r(W) > 0$, meaning that $W$ has rank $r$. 
Let $\mathcal{K}$ be the index set extracted by SPA. Then there exists a permutation $\pi$ of $\{1,2,\dots,r\}$ such that for all $k=1,2,\dots,r$, 
\[
\left\| {X}(:,\mathcal{K}(k)) - W(:,\pi(k))  \right\|_2
\; \leq \;
\mathcal{O}\left(  \frac{\epsilon K(W)^2}{ \sigma_r^2(W) } \right),
\]
where $\mathcal{K}(k)$ denotes the $k$th index in $\mathcal{K}$.
\end{theorem}

Note that the bounds in Theorem~\ref{th:spa} are relatively weak: the noise level has to be rather small to guarantee SPA to recover $W$ approximately.

\subsection{Model 2: Learning a latent simplex} \label{sec:model2}

A drawback of separable NMF algorithms, such as VCA and SPA, is that they assume that there is only one data point close to each column of $W$. 
Therefore, to estimate $W$, the column-wise bounded noise assumption (Assumption~\ref{ass:cwn}) is necessary; see Theorem~\ref{th:spa}.
This is a rather strong assumption, often not met in practical situations as typically many data points are affected by large amounts of noise.

The paper~\cite{bhattacharyya2020finding} rather proposes to leverage the fact that typically more than one data point are close to each column of $W$. 
This assumption, which is stronger than the pure-pixel one (requiring only a single pure-pixel), allows higher noise levels.
It is called the \emph{proximate latent points} assumption, and is defined as follows. 

\begin{assumption}[proximate latent points]  \label{ass:plp}
In Problem~\ref{prob:main}, there exists $r$ index sets, $\mathcal{K}_k$ for $k=1,2,\dots,r$, of cardinality at least $p = \delta n$ such that
\[
\| WH(:,j) - W(:,k) \|_2 \leq \frac{4 \sigma}{\delta}
\; \text{ for all } j \in \mathcal{K}_k,
\]
for some $\delta \in \left[\frac{1}{n},\frac{1}{r} \right]$ and $\sigma > 0$.
\end{assumption}

Under this assumption, instead of looking for one column of ${X}$ to represent each vertex, like in VCA and SPA, algorithms should look for $p$ of them and then estimate each vertex as the average of these $p$ data points. We will refer to such algorithms as \emph{smoothed separable NMF algorithms}.
The main contribution of this paper is to propose two new such algorithms; in Sections~\ref{sec:svca} and \ref{sec:sspa}.

This assumption is often met in the machine learning applications mentioned in \cref{sec:intro};
see the discussions in~\cite{bhattacharyya2020finding, bakshilearning}.
For high-resolution HSIs that satisfy the pure-pixel assumption,
there are typically more than one pixel close to each  endmember. This claim will be validated numerically in Section~\ref{sec:xphsu}.

\subsubsection{Algorithm for learning a latent simplex}

To solve Problem~\ref{prob:main} under Assumption~\ref{ass:plp},
\cite{bhattacharyya2020finding} proposed an algorithm similar to VCA, which we refer to as the algorithm for learning a latent simplex (ALLS).
The main difference between ALLS and VCA is the selection step. Instead of picking a single column of ${X}$, ALLS averages over $p$ columns for some
$p \in \left\{1,2,\dots,\lfloor \frac{n}{r} \rfloor \right\}$.
More precisely, ALLS picks the $p$ columns corresponding to the indices that maximize the absolute value of $u_k$;
see Algorithm~\ref{algo:LLS}.

The idea behind ALLS is to apply VCA on a smoothed data set. This smoothed data set is made of $\binom{n}{p}$ data points which are the averages of all possible combinations of $p$ data points, that is, $p$ columns of ${X}$. 
Of course, constructing this smoothed data set explicitly is not practical, since $\binom{n}{p}$ grows exponentially.
However, by the linearity of the selection step in VCA, this is not necessary: the smoothed data point that maximizes a linear function is the average of the $p$ data points that have the $p$ largest values for that function. 
Algorithm~\ref{algo:LLS} summarizes ALLS.
\algsetup{indent=2em}
\begin{algorithm}[ht!]
\caption{Algorithm for Learning a Latent Simplex (ALLS) \citep{bhattacharyya2020finding} 
\label{algo:LLS}}
\begin{algorithmic}[1]

\REQUIRE
The matrix $X \in \mathbb{R}^{m \times n}$,
the number~$r$ of columns of $W$,
the number $p$ of columns of $X$ to be averaged to obtain each column of $W$.    \vspace{0.1cm}

\ENSURE A matrix $W'$ such that $X \approx W' H$ for some $H \geq 0$.
    \medskip

\STATE Let $W' = [\;]$, $P^\bot = I_m$, $V = [\;]$.   \vspace{0.1cm}

\STATE {Let the column of $Y \in \mathbb{R}^{m \times r}$ be the top $r$ left singular vectors of $X$}. \vspace{0.1cm}

\FOR {$k=1$ : $r$}   \vspace{0.1cm}

\STATE Pick a random direction $d_k \in \mathbb{R}^{m}$ in the subspace spanned by $Y$,
e.g.,  $d_k \sim Y \mathcal{N}(0,I_r)$. \vspace{0.1cm}

\STATE  \label{alls:step5} Compute $u_k = \left(d_k^\top  P^\bot \right) X  \in \mathbb{R}^{n}$.   \vspace{0.1cm}

\STATE \label{alls:step6} Let $\mathcal{S}_k$ be the set of $p$ indices corresponding to the largest coordinates of $u_k$ in  absolute value.\vspace{0.1cm}

\STATE Let $W'(:,k)$ average of the columns of $X(:,\mathcal{S}_k)$.  \vspace{0.1cm}

\STATE {Update the projector $P^\bot$ (as in step \ref{li:updtproj} of Algorithm~\ref{algo:VCA}).}

\ENDFOR

\end{algorithmic}
\end{algorithm}

Note that ALLS with $p=1$ is equivalent to VCA.

\paragraph*{Computational cost}

 The only additional cost of ALLS compared to VCA is to average $p$ columns of $X$, which requires $r$ times $\mathcal{O}(pm)$ operations, which is negligible since $p \ll n \leq \nnz(X)$.

\subsubsection{Probabilistic robustness of ALLS}

Let us describe the assumptions needed to prove the probabilistic robustness of ALLS. 
The condition on $W$ is defined as follows:
\begin{assumption}[well-separatedness of $W$]  \label{ass:wsw}
In Problem~\ref{prob:main}, the matrix $W$ satisfies
\begin{equation} \label{eq:alpha}
\alpha(W)
 = 
\frac{\min_{k=1,2,\dots,r}
\min_x \| W(:,k) - W(:,\bar{k}) x \|_2 }{ K(W) }
 >  0,
\end{equation}
where $\bar{k} = \{1,2,\dots,r\} \backslash \{k\}$.
\end{assumption}
Assumption~\ref{ass:wsw} holds if and only if $\rank(W) = r$, in which case $\conv(W)$ is a simplex, that is, a polytope of dimension $r-1$ with $r$ vertices (hence the name of the algorithm).

The condition on the noise is as follows.
\begin{assumption}[Spectrally bounded perturbations]  \label{ass:sbp}
In Problem~\ref{prob:main},
\[
 \| N \|_2 = \sigma_{1}(N) 
\; \leq \; {\sigma}{\sqrt{n}} ,
\]
where there exists some constant $c$ such that
\begin{equation} \label{eq:sigma}
\sigma
\, \leq \,
\frac{\alpha^2 \, \sqrt{\delta}}{c \, r^9}
\min_{j} \, \|W(:,j)\|_2 ,
\end{equation}
where $\alpha = \alpha(W)$ is defined in
Assumption~\ref{ass:wsw}, and $\delta$ and $\sigma$ in
Assumption~\ref{ass:plp} (recall, $p = \delta n$ is the number of data points close to each column of $W$).
\end{assumption}

It is key to note here that the noise allowed is not column wise as in Assumption~\ref{ass:cwn}, but on the spectral norm of $N$, which is rather different.

We can now state the robustness theorem for ALLS.
 
\begin{theorem}~\citep{bhattacharyya2020finding} \label{th:alls}
Let us consider Problem~\ref{prob:main}
under Assumptions~\ref{ass:plp} (proximate latent points), \ref{ass:wsw} (well-separatedness of $W$) and \ref{ass:sbp} (spectrally bounded perturbations).
Then,
with probability at least $1 - c/r^{3/2}$,
ALLS computes a matrix $W'$ such that upon permutation of its columns,
for all $k=1,2,\dots,r$,
\begin{equation} \label{eq:errALLS}
\| W(:,k) - W'(:,k)||_2
\; \leq \;
O\left( \frac{r^4 \sigma}{\alpha \sqrt{\delta}} \right) .
\end{equation}
\end{theorem}

Note that substituting~\eqref{eq:sigma} in \eqref{eq:errALLS} gives
\[
\| W(:,k) - W'(:,k)||_2
\; \leq \;
O\left( \frac{ \alpha }{c \, r^5} \right) \min_{j} \, \|W(:,j)\|_2.
\]

\paragraph*{Implications for VCA} Interestingly, since ALLS for \mbox{$p=1$}  coincides  with VCA, Theorem~\ref{th:alls} provides a probabilistic robustness result for VCA which is unknown in the blind HU literature.

\subsubsection{Bounds of SPA versus ALLS}\label{subsubsec:boundsspaalls}

 Theorem~\ref{th:alls} might look somewhat weak because of the dependence in $r^9$ in the bound~\eqref{eq:sigma} for $\sigma$. However, it is not known whether this bound is tight, although it is believed it could be improved~\citep{bhattacharyya2020finding, bakshilearning}. A similar comment applies to SPA. Moreover, these bounds  assume an adversarial setting, and noise robustness under particular generative models is also an interesting direction of research,  as in~\cite{wu2021probabilistic}. 
 
 In any case, Theorem~\ref{th:alls} only requires a bound on $\|N\|_2$ while SPA requires each column of the noise matrix $N$ to be bounded, indicating that ALLS should perform better, in general, 
 when $p$ is sufficiently large.  
 Since the theory is still not fully developed and the tightness of the theoretical bounds should be carefully studied, it is important to compare these algorithms empirically to shed light on their  differences on practical problems; this will be done in \cref{sec:nex}.

\section{Smoothed VCA}  \label{sec:svca}

Inspired by VCA, and ALLS, we now propose
smoothed VCA (SVCA);
see Algorithm~\ref{algo:SVCA}.
SVCA has two key important differences compared to ALLS:

\begin{enumerate}

\item At step $k$, ALLS selects the $p$ entries maximizing the absolute value of the vector of $u_k$, obtained as the inner product of $X$ and a randomly 
generated direction $d_k^\top P^\bot$; see steps~\ref{alls:step5}-\ref{alls:step6} of Algorithm~\ref{algo:LLS}.   
This is not equivalent to maximizing (or minimizing) the  linear function $l(x) = d_k^\top P^\bot x$ over the smoothed polytope. In fact, by using the absolute value, this approach could select data points in opposite directions. For example, take the simple case with two vertices $w_1 = (-1,0)$ {and $w_2 = (1,0)$}. 
For any direction $d$, we have $|d^\top w_1| = |d^\top w_2|$ and hence it is very likely that data points close to both vertices will maximize $| d^\top x |$, and their average will be a poor approximation of both vertices.

Instead, to maximize (or minimize) $l(x)$, one should select the $p$ indices maximizing $u_k$ 
(or $-u_k$). 
In SVCA, we therefore propose to select the $p$ indices that maximize (resp.\ minimize) $u_k$ if
the median of the $p$ largest values is larger (resp.\ smaller) than the absolute value of the median of the $p$ smallest values of $u_k$.
We have observed in practical experiments that this modification of ALLS is crucial to obtain competitive results in real-world hyperspectral images. In fact, we will see that SVCA outperforms ALLS, and the main reason is this modified selection step, see lines~\ref{li:selection-if} to~\ref{li:selection-endif} of Algorithm~\ref{algo:SVCA}.

\item Instead of averaging $p$ columns of $X$ at each step, we will also consider taking their median. This allows SVCA to be much more tolerant to \revise{non-Gaussian noise, such as} gross corruptions and outliers, which are often present in HSI. \revise{ While in the presence of Gaussian noise, using the average might in principle lead to the best unmixing results, the choice of a different aggregation function might still be appealing on real world data sets, in which the linear mixing model is not always perfectly accurate. In this context, other aggregation functions might be more robust to the model imperfections (this is for instance the case in HSI, where the data sets sometimes exhibit spectral variabilities -- an aggregation function enabling to reduce the pure pixel variabilities impact would be appealing).
Using the median} also allows SVCA to be more tolerant to a misspecified value of $p$.
For example, assume a scenario where there are exactly $p'$ data points close to each vertex.
For $p < p'$, one does not leverage optimally the presence of multiple pure data points. 
On the other side, as soon as $p$ is larger than $p'$, ALLS will perform rather badly because it will average $p'$ data points close to a vertex and $p-p'$ data points potentially far away. 
If instead one takes the median, the algorithm remains able to extract the vertices accurately for any $p < 2p'$. 
In practice, as we will show in \cref{sec:nex}, using the median performs significantly better on real data sets.

\end{enumerate}

Note that SVCA has the same computational cost as VCA, SPA and ALLS, namely $\mathcal{O}(r \text{nnz}(X))$ operations.

\begin{spacing}{1.2}
\algsetup{indent=2em}
\begin{algorithm}[ht!]
\caption{Smoothed Vertex Component Analysis (SVCA)\label{algo:SVCA}}
\begin{algorithmic}[1]

\REQUIRE
The matrix $X \in \mathbb{R}^{m \times n}$,
the number~$r$ of columns of $W \in \mathbb{R}^{m \times r}$,
the number $p$ of columns of $X$ to be averaged to obtain each column of $W$, the aggregation method (median or mean).    \vspace{0.1cm}

\ENSURE A matrix $W$ such that $X \approx W H$ for some $H \geq 0$.
    \medskip

\STATE Let $W = [\;]$, $P^\bot = I_m$, $V = [ \; ]$.  \vspace{0.1cm}

\STATE Let $Y \in \mathbb{R}^{m \times r}$ be the vector space spanned by the top $r$ left singular vectors of $X$. \vspace{0.1cm}

\FOR {$k=1$ : $r$}   \vspace{0.1cm}

\STATE Pick a random direction $d_k \in \mathbb{R}^{m}$ in the subspace spanned by $Y$,
e.g.,  $d_k \sim Y \mathcal{N}(0,I_r)$. \vspace{0.1cm}

\STATE Compute $u_k = \left(d_k^\top  P^\bot \right) X  \in \mathbb{R}^{n}$.   \vspace{0.1cm}

\IF {the median of the $p$ largest values of $u_k$ is larger than the absolute value of the median of the $p$ smallest values of $u_k$ } \label{li:selection-if} \vspace{0.1cm}

\STATE Let $\mathcal{S}_k$ be the set of $p$ indices maximizing $u_k$. \vspace{0.1cm}

\ELSE {}

\STATE Let $\mathcal{S}_k$ be the set of $p$ indices minimizing $u_k$. \vspace{0.1cm}

\ENDIF \label{li:selection-endif}

\STATE Let $W(:,k)$ be the median (or the mean) of the columns of $X(:,\mathcal{S}_k)$.  \vspace{0.1cm}

\STATE {Update the projector $P^\bot$ (as in step \ref{li:updtproj} of Algorithm~\ref{algo:VCA}).}

\ENDFOR

\end{algorithmic}
\end{algorithm}
\end{spacing}

\paragraph*{Recovery guarantees for SVCA} 

SVCA is very similar to ALLS, and 
in fact the robustness analysis of ALLS applies to SVCA,  that is, Theorem~\ref{th:alls} applies to SVCA. 
The reason is that we guarantee SVCA to extract the data point in the smoothed data set that maximizes the absolute value $l(x) = d_k^\top P^\bot x$.

\section{Smoothed SPA}    \label{sec:sspa}

Since SVCA is equivalent to VCA for $p=1$, it is not guaranteed to be
deterministically robust to noise.
This motivates us to propose smoothed SPA\revise{; see Algorithm~\ref{algo:SSPA}}. 
Unfortunately, it is not practical to apply SPA directly on the smoothed data set.
Indeed, it would require to find the $p$ columns of the smoothed data set with the largest $\ell_2$ norm.
The $\ell_2$ norm being a nonlinear function, it would require to explicitly compute the $\binom{n}{p}$ data points of the smoothed data set, which is computationally prohibitive.

Instead, we replace the random selection
of $u_k = Y\mathcal{N}(0,1)$ in SVCA by the \revise{selection of the} column of the residual $P^\bot X$ with maximum $\ell_2$ norm, that is,
$u_k = P^\bot X(:,j_k)$ for some $j_k$
so that $\|u_k\|_2 \geq \|P^\bot X(:,j)\|_{2}$ for all $j$.
This allows us to combine the best of 'both worlds': deterministic robustness under separability when $p=1$, and the use of the  proximate latent point assumption when $p > 1$  (Assumption~\ref{ass:plp}). 

\begin{spacing}{1.2}
\algsetup{indent=2em}
\begin{algorithm}[ht!]
\caption{Smoothed Successive Projection Algorithm (SSPA) \label{algo:SSPA}}
\begin{algorithmic}[1]

\REQUIRE
The matrix $X \in \mathbb{R}^{m \times n}$,
the number~$r$ of columns of $W \in \mathbb{R}^{m \times r}$,
the number $p$ of columns of $X$ to be averaged to obtain each column of $W$, the aggregation method (median or mean).   

\ENSURE A matrix $W$ such that $X \approx W H$ for some $H \geq 0$.
    \medskip

\STATE Let $W = [\;]$, $P^\bot = I_m$, $V = [ \; ]$.  

\STATE Let $u_1(j) = \| X(:,j) \|_2^2$ for all $j$.

\FOR {$k=1$ : $r$}

 \STATE Let $j_k = \argmax_{1 \leq j \leq n} u_k(j)$. (Break ties arbitrarily, if necessary.)

 \STATE Let $d_k = X(:,j_k)$.

\STATE Compute $u_k = (d_k^\top P^\bot) X \in \mathbb{R}^{n}$.

\IF {$\max_i u_k(i) \geq  - \min_i u_k(i)$}  

\STATE Let $\mathcal{S}_k$ be the set of $p$ indices maximizing $u_k$.

\ELSE {}

\STATE Let $\mathcal{S}_k$ be the set of $p$ indices minimizing $u_k$.  

\ENDIF

\STATE Let $W(:,k)$ be the median (or the mean) of the columns of $X(:,\mathcal{S}_k)$.  

\STATE {Update the projector $P^\bot$ (as in step \ref{li:updtproj} of Algorithm~\ref{algo:VCA}).}

\STATE  Update the squared norms of the columns of $P^\bot X$:
 for all $j$,
\[
u_{k+1}(j) = u_{k}(j) - v_k^\top X(:,j) = \|P^\bot X(:,j)\|_2^2.
\]

\ENDFOR

\end{algorithmic}
\end{algorithm}
\end{spacing}

\paragraph*{Recovery guarantees for SSPA} 

For $p=1$, SSPA coincides with SPA, and hence Theorem~\ref{th:spa} applies to SSPA for $p=1$, that is, it is deterministically robust to column-wise bounded noise. 
However, since the selection step of SSPA is deterministic, Theorem~\ref{th:alls} does not apply to SSPA.
A promising direction of further research would be to analyze noise robustness of SSPA for $p > 1$.

\paragraph*{Should you use SVCA or SSPA?} 

SVCA has the advantage to be a randomized algorithm, and hence can be run multiple times and the best solution, according to some criterion, can be kept.
SSPA is deterministic and has the advantage to have stronger theoretical guarantees for $p=1$.   
From a practical point of view,
one could run SVCA several times, and SSPA once, 
and then keep the best solution.

\paragraph*{Comparing algorithms} 

In this paper, when the ground truth $W^*$ is not available, we will use the following criterion
\[
Q_F(W) \, = \, \frac{\min_{H \geq 0} \| X - WH \|_F}{ \| X  \|_F} \; \in \; [0,1], 
\] 
to evaluate the quality of a solution $W$.
Note that we do not use the sum-to-one constraint, $H^\top e=e$,  because, in many practical situations, including hyperspectral imaging, this constraint is not satisfied for all columns of $H$, e.g., for pixels with low luminosity; see a discussion in~\cite{gillis2014successive}.

In matrix factorization, it could be argued that in general $Q_F(W)$ is not a good measure to assess the quality of a solution $W$, since any $W$ such that $\conv({X}) \subset \conv({W})$ implies $Q_F(W) = 0$, even if $W$ is very different from the ground truth $W^*$.
In our case though, $Q_F(W)$ is relevant, since all the considered algorithms generate solutions $W$ which columns are close to $\conv({X})$.
Indeed, VCA, SPA and ALLS generate solutions within $\conv({X})$.
This is not always the case for SVCA and SSPA, because of the use of the median (a non-linear operator) for the aggregation of the columns of $X$.
In practice though, they find solutions close to $\conv({X})$.

\section{Numerical experiments} \label{sec:nex}

In this section, we study and compare the performance of ALLS, SVCA, and SSPA.
We first consider synthetic data sets\revise{.
Then we tackle} the unmixing of real-world hyperspectral images \revise{and the extraction of features in facial images data sets}.
The code and data are available online\footnote{\url{https://gitlab.com/nnadisic/smoothed-separable-nmf}}. 
All algorithms are implemented in Matlab and run on a computer with an i5-8350U processor.

\subsection{Synthetic data sets}\label{sec:xpsynth}

In this section, we study the behavior of smooth separable NMF algorithms in several experimental setups.
To build synthetic data sets, we first build $W \in \mathbb{R}^{224 \times 10}_{+}$ by selecting 10 columns from the USGS hyperspectral library\footnote{\url{https://www.usgs.gov}} using SPA. 
The condition number of the corresponding matrix is $\kappa(W) = 33.88$.
Then, we generate a random $H \in \mathbb{R}^{10 \times 1000}_{+}$ such that $H = [I_{10} , H']$, meaning there is at least one pure data point for every vertex. 
The coefficients of $H'$ follow a Dirichlet distribution, which is usually a good model for the abundances in HSI \citep{nascimento2011hyperspectral}, of parameters $\alpha e$, where $\alpha$  controls the proportion of data points close to the vertices,  
see Table~\ref{tab:alphadelta}. 
The larger $\alpha$, the denser the columns of $H$ and the less likely the `proximal latent points' assumption is to be satisfied for large $p$.
\begin{table}[ht]
\centering
\caption{Generating the columns of $H \in \mathbb{R}^{10 \times n}$ using the Dirichlet distribution of parameter $\alpha e$, this table reports the expected percentage, $\delta$, of pure data points close to each vertex. 
The  $j$th data point is considered close to the $i$th vertex  when $H(i,j) > 0.95$, hence this table reports the expected value of $\frac{1}{n} \big| \{ j \ | \ H(i,j) > 0.95 \} \big|$ for all $i$.
Since the Dirichlet distribution has the same parameters for all $i$, namely $\alpha$, this expected value is the same for all $i$.} 
\label{tab:alphadelta}
\begin{tabular}{c|c|c|c|c|c|c}
$\alpha$ & 0.01 & 0.02  & 0.05 & 0.1    & 0.2   & 0.5 \\ \hline
$\delta$ & 7.7\% & 5.9\% & 2.7\% & 0.75\% & 0.06\% & 0\%
\end{tabular}
\end{table}

Finally, we let $X = W H + N$ where $N$ is a normalized Gaussian noise: 
Given a noise level $\epsilon$, we first generate $N(i,j) \sim \mathcal{N}(0,1)$ for all $(i,j)$, then set
\[
N \leftarrow \epsilon \frac{\| WH \|_F}{\| N \|_F} N ,
\]
so that $\epsilon$ is the norm of the noise relative to $WH$:
$\| N \|_F = \epsilon \| WH \|_F$.

Given the noisy data matrix $X$ and a parameter $p$, we can compute $W'$ with the algorithms ALLS, SVCA, and SSPA. 
We note ALLS($p$), SVCA($p$) and SSPA($p$) these algorithms run with parameter $p$.
Given the computed solution $W'$, we report the mean removed spectral angle (MRSA) to assess its quality.
Given two spectral signatures $x,y \in \mathbb{R}^m$, the MRSA is defined as follows:
\[
\phi(x,y) = \frac{1}{\pi} \arccos{ \left( \frac{ (x - \bar{x})^\top (y - \bar{y}) }{ \|x - \bar{x}\|_2 \|y - \bar{y}\|_2 } \right) } \in [ 0 , 1 ] ,
\]
where for a vector $z \in \mathbb{R}^m$, $\bar{z} = (\sum_{i=1}^m z_i) e$ and $e$ is the vector of all ones.
Given two matrices, here the groundtruth $W$ and the estimate $W'$, we define the MRSA as
\[
\text{MRSA}(W, W') = \sum_{j=1}^r \phi(W(:,j) , W'(:,j)) ,
\]
after the columns of $W'$ have been reordered so as to minimize the MRSA.
The smaller the MRSA, the better the solution.
For ALLS and SVCA, on a given data set, we run 30 trials and keep the median of the results.
SSPA is deterministic so we only run it once.
Unless stated otherwise, SVCA and SSPA are equipped with the median aggregation.
In the following, we consider several experimental setups to highlight the property of these algorithms.

\begin{figure}[t]
    \centering

\begin{tikzpicture}
    \begin{axis}[
        width=\myfigsize\textwidth,
        scale only axis,
        xmode=log,
        ymode=log,
        xlabel={Noise $\epsilon$},
        ylabel={MRSA},
        legend style={at={(0.98,0.02)}, anchor=south east, font=\footnotesize}]
\addplot[color=myblue, mark=o] table [x index=0, y index=1] {xp1alls.dat}; \addlegendentry{ALLS(1)};
    
    \addplot[color=mypurple, mark=x] table [x index=0, y index=2] {xp1alls.dat}; \addlegendentry{ALLS(20)};
    \addplot[color=mypurple, mark=square] table [x index=0, y index=3] {xp1alls.dat}; \addlegendentry{ALLS(50)};
    
    \addplot[color=mygreen, mark=x] table [x index=0, y index=2] {xp1svca.dat}; \addlegendentry{SVCA(20)};
    \addplot[color=mygreen, mark=square] table [x index=0, y index=3] {xp1svca.dat}; \addlegendentry{SVCA(50)};
    
    \addplot[color=myorange, mark=o] table [x index=0, y index=1] {xp1sspa.dat}; \addlegendentry{SSPA(1)=SPA};
    \addplot[color=myorange, mark=x] table [x index=0, y index=2] {xp1sspa.dat}; \addlegendentry{SSPA(20)};
    \addplot[color=myorange, mark=square] table [x index=0, y index=3] {xp1sspa.dat}; \addlegendentry{SSPA(50)};
    \end{axis}
\end{tikzpicture}

     \caption{Results for ALLS, SVCA, and SSPA for different values of $p$, when $\epsilon$ varies, for fixed $n=1000$ and purity $\alpha=0.05$ ($\delta=2.7\%$). Values for ALLS and SVCA are the medians over 30 trials. Note that ALLS(1)=SVCA(1)=VCA.
    \label{fig:xpsynth1}}
\end{figure}
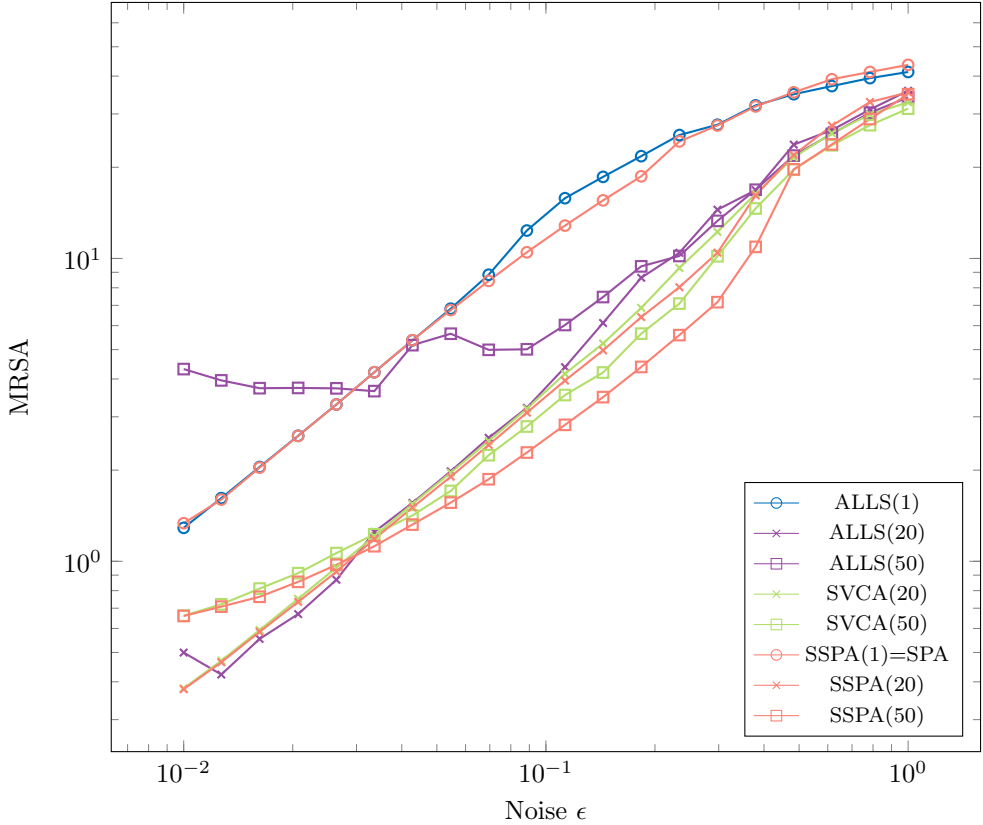

In \cref{fig:xpsynth1}, we test ALLS, SVCA, and SSPA with different values of the parameter $p$, when the noise $\epsilon$ varies.
Note that SVCA(1) and SSPA(1) are equivalent to their non-smoothed version VCA and SPA.
Also note that ALLS(1) is equivalent to SVCA(1) and thus VCA.
In this experiment, we observe that smoothing improves the algorithm  performances. 
However, for ALLS, a parameter $p$ set too large can in fact worsen the solution, especially when the noise level is small.
Also, ALLS is outperformed by SVCA and SSPA; this will be confirmed in experiments on hyperspectral images in \cref{sec:xphsu}.

\begin{figure}[t]
    \centering
    
\begin{tikzpicture}
    \begin{axis}[
        width=\myfigsize\textwidth,
        scale only axis,
        xmode=log,
        ymode=log,
        xlabel={Noise $\epsilon$},
        ylabel={MRSA},
        legend style={at={(0.98,0.02)}, anchor=south east, font=\footnotesize}]
    
    \addplot[color=mypurple, mark=square] table [x index=0, y index=3] {xp6.dat}; \addlegendentry{Best ALLS};
    \addplot[color=mypurple, mark=otimes] table [x index=0, y index=2] {xp6.dat}; \addlegendentry{Median ALLS};
    \addplot[color=mypurple, mark=triangle] table [x index=0, y index=1] {xp6.dat}; \addlegendentry{Worst ALLS};
    
    \addplot[color=mygreen, mark=square] table [x index=0, y index=6] {xp6.dat}; \addlegendentry{Best SVCA};
    \addplot[color=mygreen, mark=otimes] table [x index=0, y index=5] {xp6.dat}; \addlegendentry{Median SVCA};
    \addplot[color=mygreen, mark=triangle] table [x index=0, y index=4] {xp6.dat}; \addlegendentry{Worst SVCA};

    \addplot[color=blue, mark=diamond] table [x index=0, y index=8] {xp6.dat}; \addlegendentry{Best err SVCA};

    \addplot[color=myorange, mark=o] table [x index=0, y index=7] {xp6.dat}; \addlegendentry{SSPA};

    \end{axis}

\end{tikzpicture}

     \caption{Comparison of SSPA with best, median, and worst result for ALLS and SVCA among 30 runs, when $\epsilon$ varies, for fixed $n=1000$, purity $\alpha=0.05$ ($\delta=2.7\%$) and parameter $p=25$.
    "Best err SVCA" represents the MRSA of the solution of SVCA that has the smallest relative reconstruction error.
    \label{fig:xpsynth6}}
\end{figure}
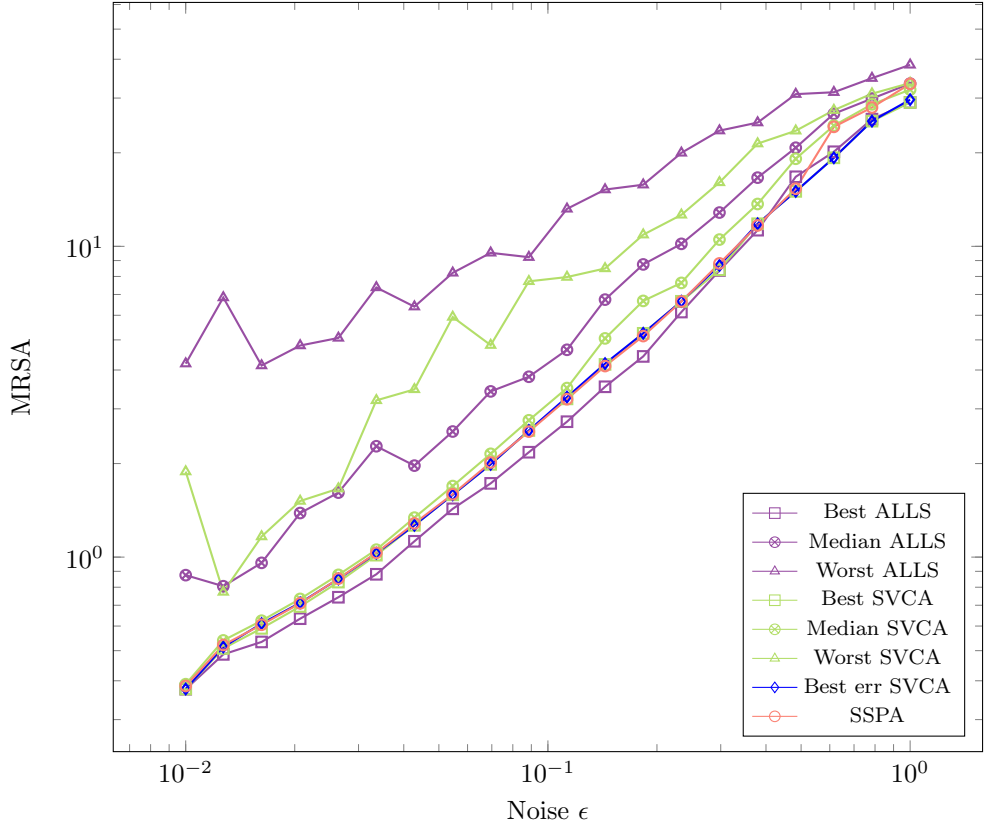

\begin{figure}[t]
    \centering

\begin{tikzpicture}
    \begin{axis}[
        width=\myfigsize\textwidth,
        scale only axis,
ymode=log,
        xlabel={Parameter $p$},
        ylabel={MRSA},
        legend style={at={(0.2,0.97)}, anchor=north west, font=\footnotesize}]
    
    \addplot[color=mygreen, mark=square] table [x index=0, y index=1] {xp4.dat}; \addlegendentry{SVCA-med};
    \addplot[color=mygreen, mark=triangle, dashed] table [x index=0, y index=2] {xp4.dat}; \addlegendentry{SVCA-mean};
    
    \addplot[color=myorange, mark=square] table [x index=0, y index=3] {xp4.dat}; \addlegendentry{SSPA-med};
    \addplot[color=myorange, mark=triangle, dashed] table [x index=0, y index=4] {xp4.dat}; \addlegendentry{SSPA-mean};
    
    \end{axis}

\end{tikzpicture}

     \caption{Results for SVCA and SSPA using either the median or the mean to average points, when $p$ varies, for fixed $n=1000$, purity $\alpha=0.05$ ($\delta=2.7\%$), and noise $\epsilon=0.05$. Values for SVCA are the medians over 30 trials.
    \label{fig:xpsynth4}}
\end{figure}
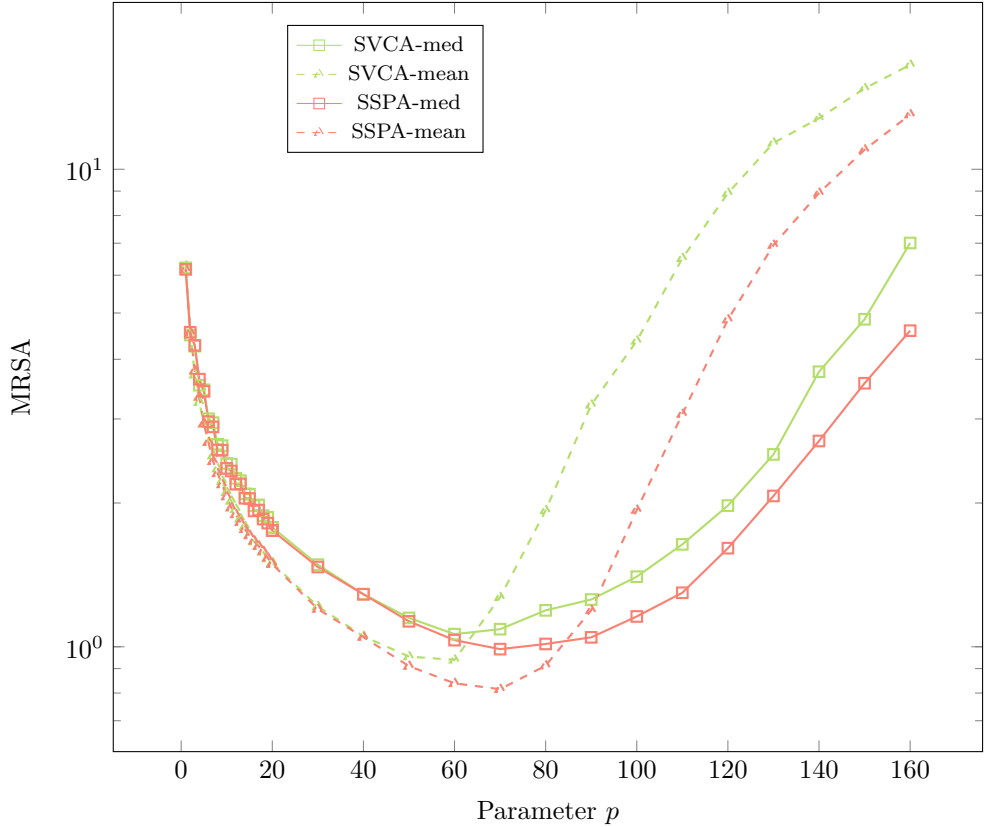

In \cref{fig:xpsynth6}, we compare the stability of ALLS and SVCA when $\epsilon$ varies by showing the best, median, and worst result among 30 runs.
We also compare them to the {MRSA of the result of SVCA that has the smallest reconstruction error}, $Q_F(W')$, and to SSPA \revise{which is deterministic}. 
We see that the best result from ALLS is slightly better than other results. 
This is due to the use of the mean as an aggregation method, which works better with the centered Gaussian noise of the synthetic data\footnote{Using the mean aggregation instead of the median aggregation in SVCA, its best MRSA result is always better than the best MRSA of ALLS in this experiment.}, see \cref{fig:xpsynth4}. 
However, the algorithm is less stable, as \revise{its} median and worst results are worst than SVCA.
With SVCA, the median results are close to the best. 
Also, the best results in terms of reconstruction error generally coincides with the best one in terms of MRSA, showing that the reconstruction error is a good proxy for the tested algorithms, as evoked in \cref{sec:sspa} (this will be useful when the groundtruth is unknown and the MRSA cannot be computed, for example in \cref{sec:xphsu}). 
The deterministic \revise{algorithm SSPA produces results that are slightly better than the} median result of SVCA. \revise{However, the best result from SVCA is better than the result from SSPA for most noise levels.}

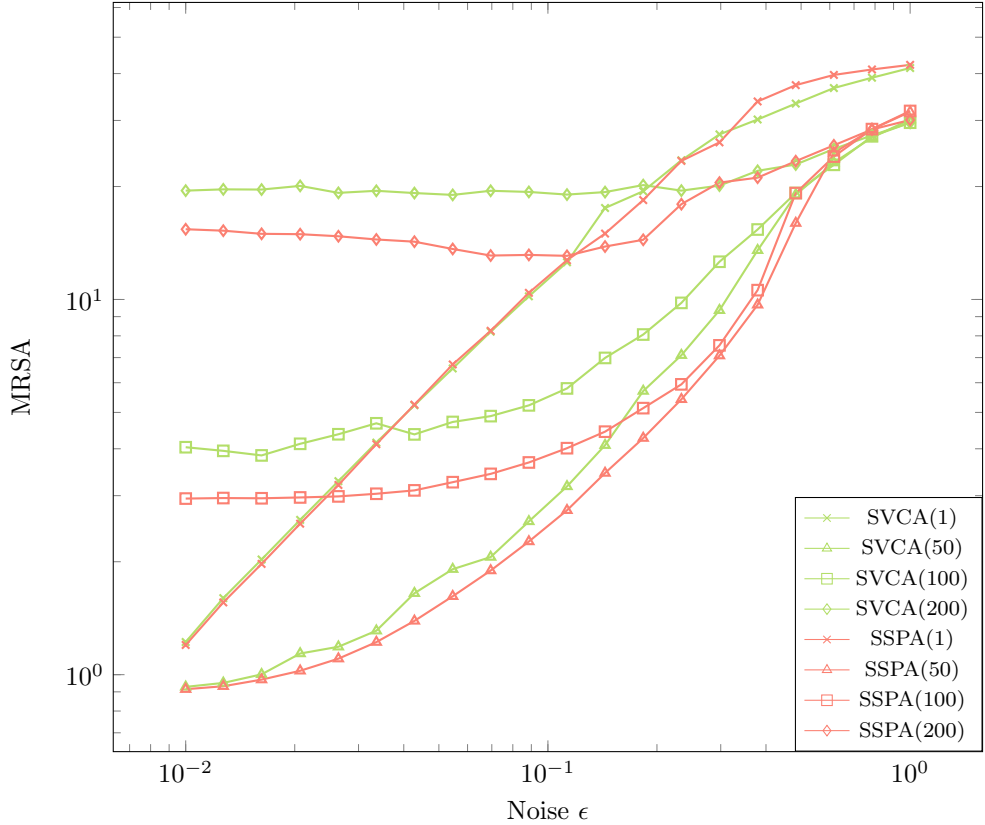
\begin{figure}[t]
    \centering

\begin{tikzpicture}
    \begin{axis}[
        width=\myfigsize\textwidth,
        scale only axis,
        xmode=log,
        ymode=log,
        xlabel={Noise $\epsilon$},
        ylabel={MRSA},
        legend style={at={(1,0.0)}, anchor=south east, font=\footnotesize}]
    
    \addplot[color=mygreen, mark=x] table [x index=0, y index=1] {xp3.dat}; \addlegendentry{SVCA(1)};
    \addplot[color=mygreen, mark=triangle] table [x index=0, y index=2] {xp3.dat}; \addlegendentry{SVCA(50)};
    \addplot[color=mygreen, mark=square] table [x index=0, y index=3] {xp3.dat}; \addlegendentry{SVCA(100)};
    \addplot[color=mygreen, mark=diamond] table [x index=0, y index=4] {xp3.dat}; \addlegendentry{SVCA(200)};
    
    \addplot[color=myorange, mark=x] table [x index=0, y index=5] {xp3.dat}; \addlegendentry{SSPA(1)};
    \addplot[color=myorange, mark=triangle] table [x index=0, y index=6] {xp3.dat}; \addlegendentry{SSPA(50)};
    \addplot[color=myorange, mark=square] table [x index=0, y index=7] {xp3.dat}; \addlegendentry{SSPA(100)};
    \addplot[color=myorange, mark=diamond] table [x index=0, y index=8] {xp3.dat}; \addlegendentry{SSPA(200)};
    
    \end{axis}

\end{tikzpicture}

     \caption{Results for SVCA and SSPA for different values of $p$, when $\epsilon$ varies, for fixed $n=1000$ and purity $\alpha=0.05$ ($\delta=2.7\%$), Values for SVCA are the medians over 30 trials.
    \label{fig:xpsynth3}}
\end{figure}

In \cref{fig:xpsynth3}, we compare SVCA and SSPA with higher values of $p$ when $\epsilon$ varies.
Again, we observe that the smoothing improves the algorithm performances, but an overestimated $p$ worsens it.
Interestingly, when the noise is very high (above 10\%), smoothed algorithms outperform their non-smoothed counterpart even when $p$ is overestimated. This is due to the fact that the value of $p$ required to obtain the best estimation of $W$ is not only determined by the purity but also by the noise level.
For instance, consider a toy example with four data points and $r = 2$: 
\begin{equation*}
    X = WH + N = W
    \begin{bmatrix}
    1\ 0\ 0.99\ 0.01\\
    0\ 1\ 0.01\ 0.99
    \end{bmatrix}
    + N.
\end{equation*}
That is, $x_3$ and $x_4$ are not pure-pixel but are almost pure. Let us further assume $N$ to follow a centered Gaussian law. If $\epsilon = 0$ (noiseless mixing), the best estimation of $W(:,1)$ (resp. $W(:,2)$) from $X$ is to extract $X(:,1)$ (resp. $X(:,2)$), yielding a perfect estimation. On the other hand, if $\epsilon$ is large relatively to the distance of $W(:,1)$ and $W(:,2)$, it is better to choose as an estimate of $W(:,1)$ (resp. $W(:,2)$) the average -- or median -- of $X(:,1)$ and $X(:,3)$ (resp. $X(:,2)$ and $X(:,4)$), as the noise power is then divided by two.

In \cref{fig:xpsynth4}, we compare SVCA and SSPA equipped with either the median or the mean aggregation, for fixed data setup and when $p$ varies.
The reverse bell curve shows that the performance of the algorithms improves gradually as $p$ grows, until it reaches an optimal value, after which the performance gradually worsens.
We observe that the algorithms equipped with the median are more robust to an overestimation of $p$, but with the mean they are slightly better for smaller $p$.
However the difference is small; this is expected as this synthetic data is generated with centered Gaussian noise, and as such the mean is expected to give the best estimation when $p$ is well chosen.
The difference is more obvious in hyperspectral images, \revise{for which the median aggregation always performs better than the mean aggregation}; see \cref{sec:xphsu,fig:xphsu2}.

\revise{
In \cref{fig:xppoisson}, we compare SVCA and SSPA equipped with either the median or the mean aggregation, for fixed data setup and when $p$ varies, but where the synthetic data has been generated using the  Poisson distribution instead of Gaussian noise. 
This data generation accounts for a counting process~\cite{hasinoff2014photon}. 
Given a noise level $\epsilon$, we first generate random matrices $W$ and $H$ as above.
Then we compute $c = \frac{1}{\text{mean}(WH) \epsilon^2}$, where $\text{mean}(WH)$ is the average of the entries of $WH$, and use the Matlab function \texttt{poissrnd} to generate a noisy data matrix $X = \frac{1}{c} {\texttt{poissrnd}(cWH)}$, that is, the entry $X_{i,j}$ of $X$ follows a Poisson distribution of parameter $c(WH)_{i,j}$, divided by $c$. This choice of $c$ allows one to control the norm of the noise relative to the groundtruth $WH$ (recall that the expectation of the Poisson distribution of parameter $\lambda$ is $\lambda$, and the variance is $\lambda$). 
The expected value of the entries of $X$ remains unchanged (namely, $\text{mean}(WH)$), while the standard deviation is equal to $\epsilon \ \text{mean}(WH)$. Hence, the entries of $X$ are close to the interval $[WH-\epsilon, WH+\epsilon]$, and we have that $\|X-WH\|_F/\|WH\|_F \approx \epsilon$.  
We observe again that the smoothed algorithms outperform significantly VCA and SPA.
A notable difference compared to experiments with Gaussian noise is that, with Poisson noise, the algorithms equipped with the median aggregation give, in general, better results than with the mean.
The median aggregation is still more robust to an overestimation of $p$.
As opposed to previous experiments, the median results of SVCA are better than the results of SSPA.
This is expected because the selection step of SSPA uses the $\ell_2$-norm, which makes it more sensitive to non-Gaussian noise.

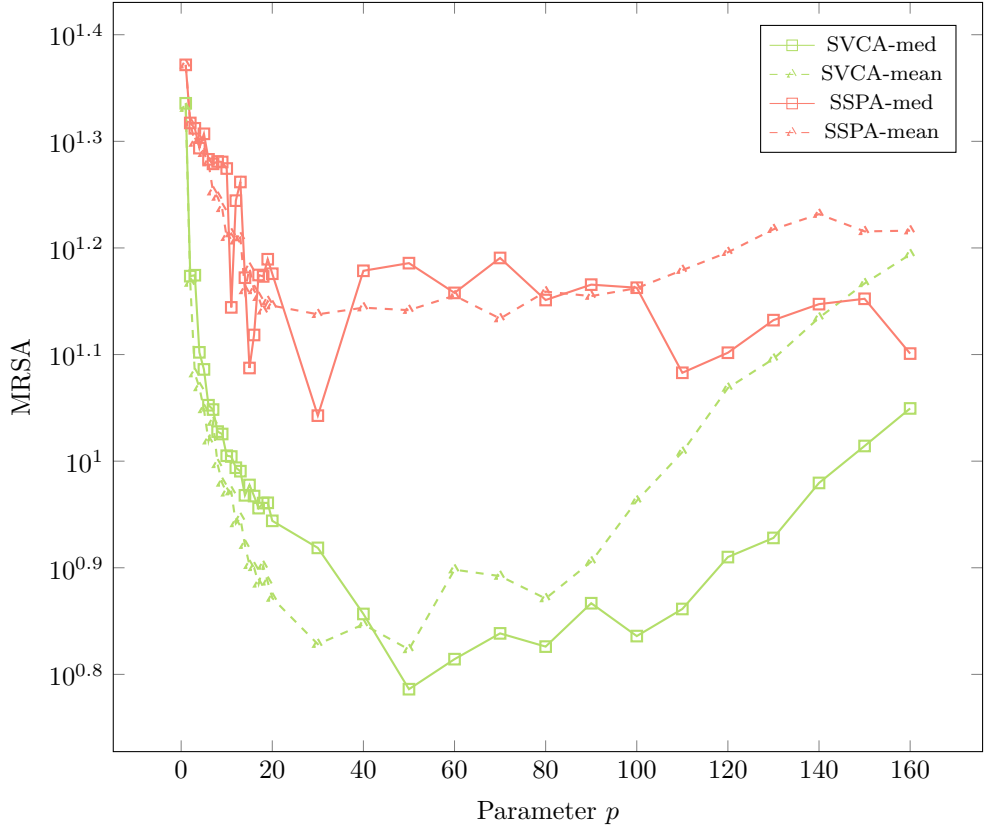
\begin{figure}[t]
    \centering

\begin{tikzpicture}
    \begin{axis}[
        width=\myfigsize\textwidth,
        scale only axis,
ymode=log,
        xlabel={Parameter $p$},
        ylabel={MRSA},
legend style={at={(0.97,0.97)}, anchor=north east, font=\footnotesize}]
    
    \addplot[color=mygreen, mark=square] table [x index=0, y index=1] {xppoisson.dat}; \addlegendentry{SVCA-med};
    \addplot[color=mygreen, mark=triangle, dashed] table [x index=0, y index=2] {xppoisson.dat}; \addlegendentry{SVCA-mean};
    
    \addplot[color=myorange, mark=square] table [x index=0, y index=3] {xppoisson.dat}; \addlegendentry{SSPA-med};
    \addplot[color=myorange, mark=triangle, dashed] table [x index=0, y index=4] {xppoisson.dat}; \addlegendentry{SSPA-mean};
    
    \end{axis}

\end{tikzpicture}

     \revise{
    \caption{Results for SVCA and SSPA using either the median or the mean to average points, when $p$ varies, for Poisson noise of level $\epsilon=0.1$, for fixed $n=1000$, purity $\alpha=0.02$ ($\delta=5.9\%$). Values for SVCA are the medians over 30 trials.
    \label{fig:xppoisson}}}
\end{figure}
}

\begin{figure}[t]
    \centering

\begin{tikzpicture}
    \begin{axis}[
        width=\myfigsize\textwidth,
        scale only axis,
ymode=log,
        xlabel={Parameter $p$},
        ylabel={MRSA},
        legend style={at={(0.2,0.97)}, anchor=north west, font=\footnotesize}]
    
    \addplot[color=mygreen, mark=square] table [x index=0, y index=1] {xp5.dat}; \addlegendentry{SVCA $\alpha$=0.01};
    \addplot[color=mygreen, mark=otimes] table [x index=0, y index=2] {xp5.dat}; \addlegendentry{SVCA $\alpha$=0.02};
    \addplot[color=mygreen, mark=triangle] table [x index=0, y index=3] {xp5.dat}; \addlegendentry{SVCA $\alpha$=0.05};

    \addplot[color=myorange, mark=square] table [x index=0, y index=4] {xp5.dat}; \addlegendentry{SSPA $\alpha$=0.01};
    \addplot[color=myorange, mark=otimes] table [x index=0, y index=5] {xp5.dat}; \addlegendentry{SSPA $\alpha$=0.02};
    \addplot[color=myorange, mark=triangle] table [x index=0, y index=6] {xp5.dat}; \addlegendentry{SSPA $\alpha$=0.05};
    
    \end{axis}

\end{tikzpicture}

     \caption{Results for SVCA and SSPA for different values of purity $\alpha$, when $p$ varies, for fixed $n=1000$ and noise $\epsilon=0.05$. Values for SVCA are the medians over 30 trials. 
    \label{fig:xpsynth5}}
\end{figure}
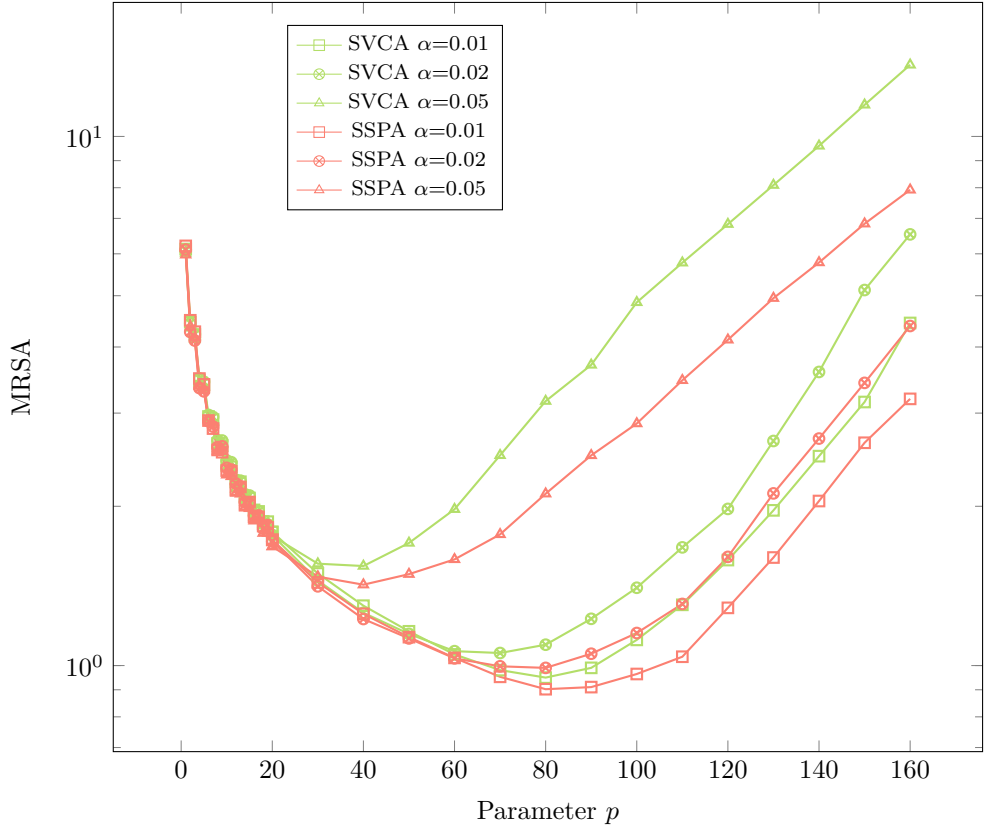

In \cref{fig:xpsynth5}, we compare SVCA and SSPA when $p$ varies in setups with different values of purity $\alpha$.
As expected, we observe that the shapes of the curves are similar, and that for a fixed noise level the value of the parameter $p$ leading to the lowest MRSA decreases as the parameter $\alpha$ increases.  

To summarize, our synthetic experiments highlight that the two proposed algorithms consistently obtain better results than ALLS. They furthermore show that SVCA and SSPA outperform VCA and SPA when $p$ is well chosen. 
Although the choice of $p$ is important, as a bad value can worsen the results compared to the non-smoothed algorithms, the use of the median \revise{aggregation} instead of the mean makes the algorithms less sensitive to this choice. 
\revise{
The use of the median is also more adapted to 
sparse, Laplace or Poisson noises, whereas the mean gives better results in the case of centered Gaussian noise.}

\subsection{Hyperspectral images}\label{sec:xphsu}

In this section, we apply ALLS, SVCA, and SSPA to the unmixing of hyperspectral images, as described in \cref{sec:intro}.
We consider three commonly used hyperspectral images\footnote{Downloaded from \url{http://lesun.weebly.com/hyperspectral-data-set.html}}, San Diego, Urban, and Terrain.
In hyperspectral data sets, extremely large values are commonly associated with sensor noise or interference. 
To avoid overfitting the factorizations to these interferences, the pixels corresponding to the 10 largest values of any wavelength range are zeroed out.
Extreme pixels generally have extreme values in many wavelenght ranges at once, so
this preprocessing results in the removal of less than 0.1\% pixels. The characteristics of these images are summarized in \cref{tab:datasets}.

\begin{table}[ht]
\caption{Summary of the hyperspectral images studied in this work.}\label{tab:datasets}
\centering
\begin{tabular}{l|l|l|l|l}
Data set   & $m$    & $n$                        & $r$  & Pixels zeroed out\\ \hline
San Diego & 158 & $400 \times 400 = 160000$ & 8 & 19 \\
Urban     & 162 & $307 \times 307 = 94249$  & 6 & 68\\
Terrain   & 188 & $500 \times 307 = 153500$ & 6 & 107\\
\end{tabular}
\end{table}

Given a data matrix $X \in \mathbb{R}^{m\times n}$, we compute $W \in \mathbb{R}^{m \times r}$ with the three algorithms.
We then compute for each algorithm $H \in \mathbb{R}^{r \times n}$ with a standard coordinate descent algorithm~\citep{gillis2012accelerated}
and measure the relative reconstruction error $\|X-WH\|_F / \|X\|_F$.
The smaller the error, the better the solution.

Some works such as \cite{zhu2017hyperspectral} proposed groundtruths for these hyperspectral images, but they are computed using numerical methods and as such do not necessarily represent reality.
Therefore, we lack a reference to assess the quality of the reconstruction, for example by measuring the MRSA.
This is why we use the relative reconstruction error as the criterion for the quality of a solution.
This is a satisfying criterion, as illustrated in \cref{fig:xpsynth6}.

In \cref{tab:xphsu}, we report results from the experiments.
We observe that, when $p>1$, the result is always improved.
When $p$ is too large, however, the solution can be worse than for $p=1$, 
as expected. 
We observe that the best $p$ varies between the algorithms.
For example, with Terrain, ALLS and SVCA perform best with $p=1000$ while SSPA performs best with $p=100$.
Larger values of $p$ seem to give more stable results, as it produces solutions with smaller deviations. 
SVCA outperforms ALLS in all cases.
SSPA performance is comparable to SVCA, and generally produces a better result than the median of SVCA, but never better than the best result obtained by SVCA.

In a few cases, SSPA produces solutions with a large error, for example in San Diego for $p=100$ and Terrain for $p=1000$.
We believe this behavior to originate from small groups of points with a very large norm, that could correspond to a rare material or to interference.

\begin{table*}[ht]
\caption{Relative reconstruction errors ($\min_{H \geq 0} \| X - WH \|_F / \| X \|_F$) resulting from the unmixing of hyperspectral images with ALLS, SVCA, and SSPA, with different values of parameter $p$. SVCA(1) and SSPA(1) are equivalent to VCA and SPA. For non-deterministic algorithms ALLS and SVCA, we show the minimum, median, standard deviation, and maximum of the error over 30 trials.}
\label{tab:xphsu}
\resizebox{\textwidth}{!}{\centering
\begin{tabular}{ll|ccc|ccc|ccc}
     &      & \multicolumn{3}{c|}{SanDiego}  & \multicolumn{3}{c|}{Urban}      & \multicolumn{3}{c}{Terrain}  \\
     & p    & Min  & Med $\pm$ std   & Max   & Min  & Med $\pm$ std    & Max   & Min  & Med $\pm$ std   & Max  \\ \hline
ALLS & 1    & 4.72 & 5.60 $\pm$ 0.67 & 8.25  & 5.39 & 9.14 $\pm$ 1.93  & 12.26 & 3.94 & 4.88 $\pm$ 0.73 & 7.08 \\
     & 100  & \textbf{4.27} & \textbf{5.35} $\pm$ 1.72 & 10.91 & \textbf{6.37} & \textbf{9.28} $\pm$ 3.18  & 19.40 & \textbf{3.84} & 4.87 $\pm$ 0.87 & 6.85 \\
     & 1000 & 4.64 & 6.14 $\pm$ 1.12 & 8.68  & 6.78 & 9.71 $\pm$ 2.16  & 14.20 & \textbf{3.84} & \textbf{4.71} $\pm$ 1.19 & 8.81 \\
     & 2000 & 4.87 & 5.91 $\pm$ 1.62 & 11.79 & 6.96 & 9.93 $\pm$ 1.60  & 12.85 & 3.96 & 4.89 $\pm$ 0.88 & 7.63 \\
     & 5000 & 5.51 & 7.42 $\pm$ 2.64 & 13.88 & 7.68 & 10.37 $\pm$ 1.89 & 14.98 & 4.28 & 5.26 $\pm$ 0.80 & 6.88 \\ \hline
SVCA & 1    & 3.95 & 5.42 $\pm$ 0.61 & 6.90  & 6.25 & 9.13 $\pm$ 1.78  & 12.23 & 4.03 & 5.11 $\pm$ 1.25 & 8.70 \\
     & 100  & \textbf{3.44} & 4.92 $\pm$ 0.77 & 6.96  & \textbf{5.08} & \textbf{6.10} $\pm$ 1.27  & 10.13 & 3.52 & 4.04 $\pm$ 0.67 & 6.52 \\
     & 1000 & 3.82 & 4.95 $\pm$ 0.59 & 6.82  & 5.82 & 6.77 $\pm$ 1.23  & 10.84 & \textbf{3.18} & \textbf{3.92} $\pm$ 0.38 & 4.70 \\
     & 2000 & 3.73 & \textbf{4.40} $\pm$ 0.51 & 5.81  & 5.66 & 6.36 $\pm$ 0.67  & 7.83  & 3.38 & 4.12 $\pm$ 0.45 & 4.95 \\
     & 5000 & 4.01 & 4.66 $\pm$ 0.73 & 7.01  & 5.69 & 6.94 $\pm$ 1.21  & 11.74 & 3.70 & 4.19 $\pm$ 0.30 & 4.82 \\ \hline
SSPA & 1    & \multicolumn{3}{c|}{5.90}      & \multicolumn{3}{c|}{9.46}       & \multicolumn{3}{c}{5.01}     \\
     & 100  & \multicolumn{3}{c|}{9.29}      & \multicolumn{3}{c|}{6.65}       & \multicolumn{3}{c}{\textbf{4.03}}     \\
     & 1000 & \multicolumn{3}{c|}{5.82}      & \multicolumn{3}{c|}{6.22}       & \multicolumn{3}{c}{8.05}     \\
     & 2000 & \multicolumn{3}{c|}{\textbf{4.32}}      & \multicolumn{3}{c|}{6.11}       & \multicolumn{3}{c}{7.86}     \\
     & 5000 & \multicolumn{3}{c|}{4.65}      & \multicolumn{3}{c|}{\textbf{5.91}}       & \multicolumn{3}{c}{5.38}   
\end{tabular}}
\end{table*}

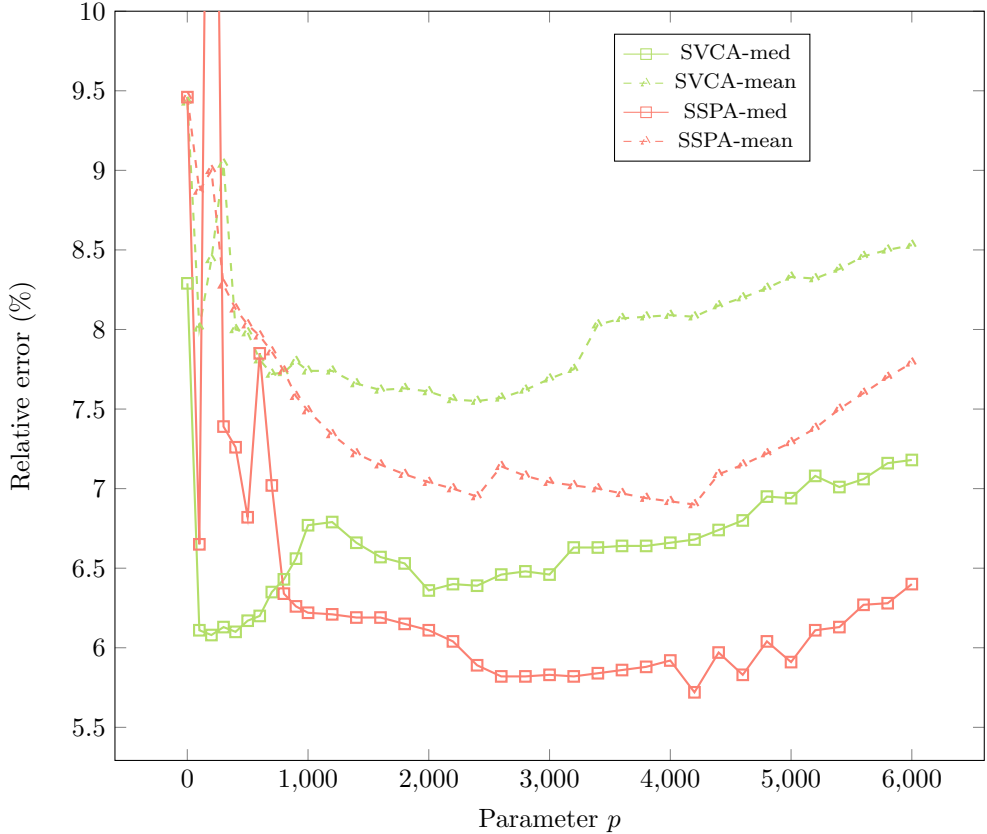
\begin{figure}[ht]
    \centering

\begin{tikzpicture}
    \begin{axis}[
        width=\myfigsize\textwidth,
        scale only axis,
xlabel={Parameter $p$},
        ylabel={\revise{Relative error (\%)}},
        ymax=10,
        legend style={at={(0.8,0.97)}, anchor=north east, font=\footnotesize}]
    
    \addplot[color=mygreen, mark=square] table [x index=0, y index=1] {xphsu2-urbanpp.dat}; \addlegendentry{SVCA-med};
    \addplot[color=mygreen, mark=triangle, dashed] table [x index=0, y index=2] {xphsu2-urbanpp.dat}; \addlegendentry{SVCA-mean};
    
    \addplot[color=myorange, mark=square] table [x index=0, y index=3] {xphsu2-urbanpp.dat}; \addlegendentry{SSPA-med};
    \addplot[color=myorange, mark=triangle, dashed] table [x index=0, y index=4] {xphsu2-urbanpp.dat}; \addlegendentry{SSPA-mean};
    
    \end{axis}

\end{tikzpicture}
     \caption{Results of the unmixing of the hyperspectral image Urban. Values for SVCA are the medians over 30 trials. One point is out of the plot; for $p=200$, SVCA-med has an error of $14.55\%$.} 
    \label{fig:xphsu2}
\end{figure}

In \cref{fig:xphsu2}, we compare SVCA and SSPA using either the median or the mean for the unmixing of the hyperspectral image Urban, with a varying $p$.
We observe that $p>1$ always leads to better results for all algorithms and all data sets. 
Also, the median aggregation almost always gives better results than the mean.
While the curves are not as regular as with synthetic data, we observe a similar tendency that the solution improves when $p$ grows, until a certain point or zone after which it worsens again.
However, SSPA-med has an irregular behaviour for $p=200$. 
This can be explained by the fact that SSPA is a greedy algorithm, so if it makes a bad choice in the first iterations, it will likely never compensate.
Also, it is deterministic, so the error is not averaged over several runs. 

\begin{figure*}[ht!]
\centering
\subfloat[VCA, error$=6.24\%$]{
    \includegraphics[width=0.98\textwidth]{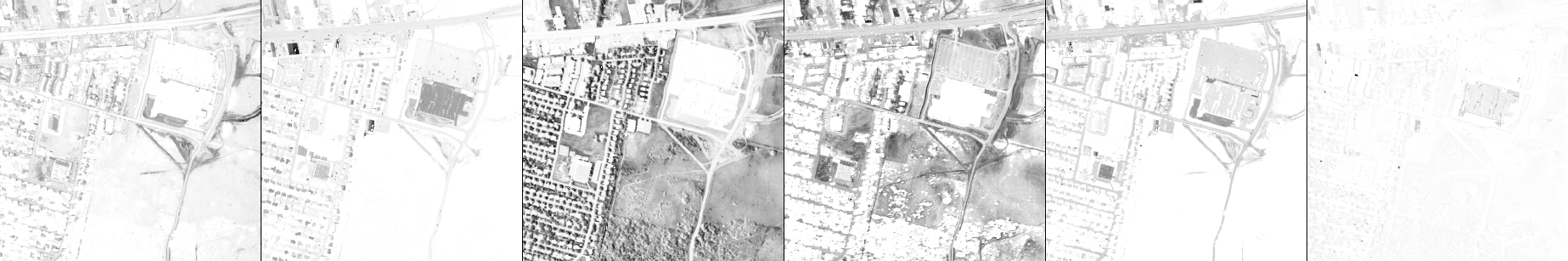}
}\\
\subfloat[SVCA $p$=200, error$=5.24\%$]{
    \includegraphics[width=0.98\textwidth]{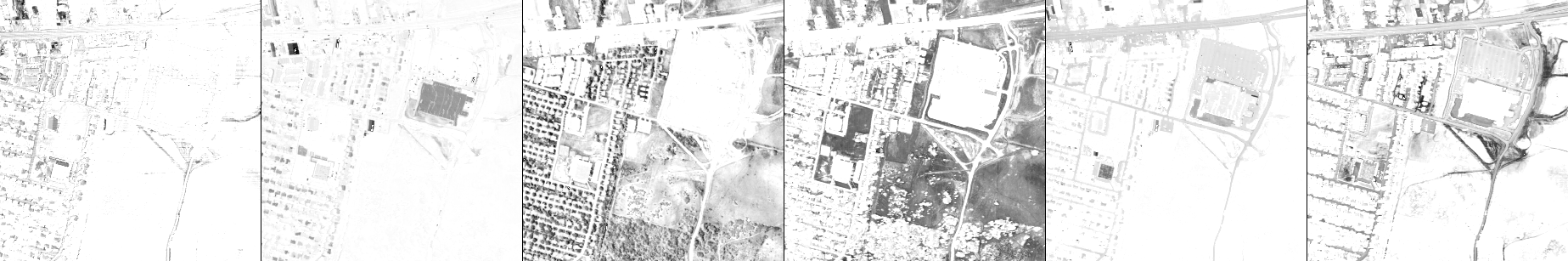}
}\\
\subfloat[SPA, error$=9.46\%$]{
    \includegraphics[width=0.98\textwidth]{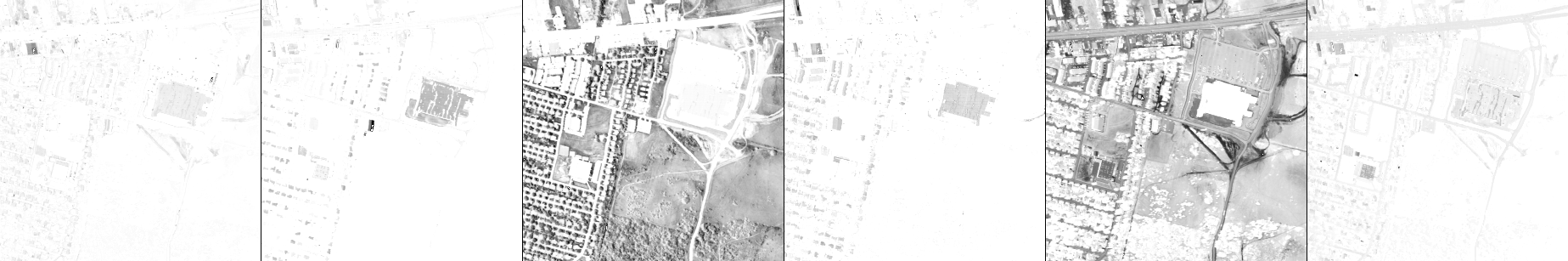}
}\\
\subfloat[SSPA $p$=4200, error$=5.72\%$]{
    \includegraphics[width=0.98\textwidth]{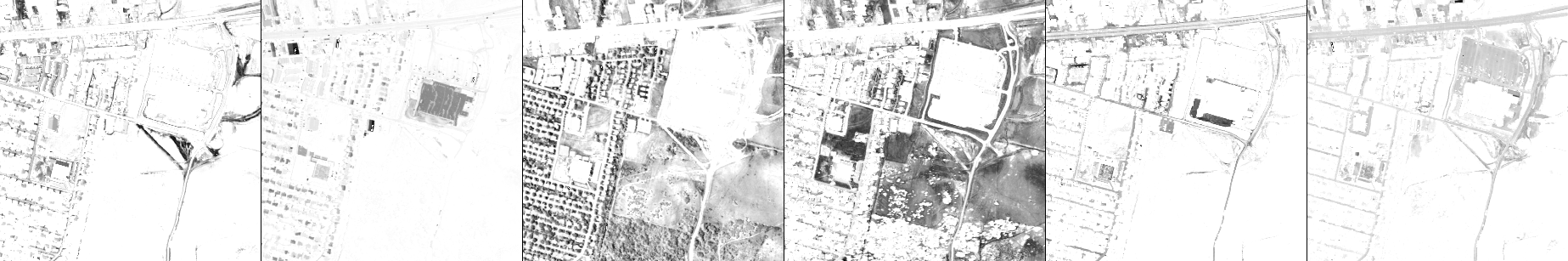}
}\\
  \caption{Abundance maps of the unmixing of the Urban hyperspectral images (that is, reshaped rows of $H$) with different algorithms. Endmembers have been reordered for easier comparison. Parameters $p$ have been chosen as the best from \cref{fig:xphsu2}. Error corresponds to $\min_{H \geq 0} \| X - WH \|_F / \| X \|_F$. For VCA and SVCA, we show the best solution over 30 trials.}
  \label{fig:abundancemaps}
\end{figure*}

In \cref{fig:abundancemaps}, we show the abundances maps corresponding to the unmixing of Urban; the corresponding spectral signatures can be found in the supplementary material (page~\pageref{morexp}).  
They indicate the proportion of every of the 6 extracted endmembers in the pixels of the image.
We see that the smoothed algorithms obtain a better separation than the non-smoothed ones.
For example, the fourth endmember extracted by SVCA and SSPA corresponds to grass, and it is well separated by these algorithms, while VCA and SPA mix it with asphalt and dirt.
The second endmember extracted by SVCA and SSPA corresponds to metallic rooftops, and it is well separated while VCA mixes it with other materials and SPA does not clearly identify it and produces a blurred picture.

In the supplementary material (page~\pageref{morexp}), we provide additional experiments, namely, the sensitivity to $p$ of SVCA and SSPA, and the abundance maps and spectral signatures of the different algorithms for the hyperspectral images Terrain and San Diego.

\revise{
\subsection{Facial images} \label{sec:xpfaces}

A popular application of NMF is the extraction of facial features, such as eyes, noses and lips, in a set of images, because it was first introduced in the seminal paper of Lee and Seung~\cite{lee1999learning}.  
In this application, each row of $X$ is a vectorized facial image, and an NMF, $WH \approx X$, extracts vectorized facial features as the rows of $H$. 
Separability makes sense in this context: it requires that, for each facial feature, there is at least one pixel that is only contained in that facial feature; see~\cite[p.~209]{gillis2020book} for a discussion and a numerical example. 
For the proximate latent points assumption to make sense, we need for each facial feature several pixels that appear mostly in that feature.
Hence the more pixels in the images, the more likely this assumption will hold and the larger the parameter $p$ can be chosen; this will be illustrated on the numerical examples below.  

Let us now apply ALLS, SVCA, and SSPA to extract facial features on four widely used data sets; see Table~\ref{tab:facedatasets}.
Note that the first two data sets (CBCL, Frey) have significantly fewer pixels than the last two (ORL, UMIST) and hence we expect that $p$ can be chosen larger for the last two. 
We choose $r=10$ for CBCL and Frey, and $r=20$ for ORL and UMIST. 
\begin{table}[ht]
\revise{
\caption{Summary of the facial images studied in this work.}\label{tab:facedatasets}
\centering
\begin{tabular}{l|l|l|l}
Data set   & $m$ ($\#$ faces)    & $n$ ($\#$ pixels)                        & $r$  \\ \hline
CBCL$^1$   & 2429 & $19 \times 19 = 361$  & 10  \\
Frey$^2$ & 1965 & $28 \times 20 = 560$ & 10  \\
ORL$^3$     & 400 & $112 \times 92 = 10304$  & 20 \\
UMIST$^2$   & 565 & $112 \times 92 = 10304$ & 20 \\
\end{tabular}
\begin{flushleft}
\footnotesize 
$^1$ \url{https://gitlab.com/ngillis/nmfbook/-/blob/master/data\%20sets/CBCL.mat} \\ 
$^2$ \url{http://www.cs.toronto.edu/~roweis/data.html}\\
$^3$ \url{https://cam-orl.co.uk/facedatabase.html} 
\end{flushleft}
}
\end{table}

\Cref{tab:xpfaces1} shows the results from the experiments of facial images data sets.
The reconstruction errors are roughly between $15\%$ and $25\%$, meaning that this type of data does not follow closely the NMF model.
However, this model is still able to extract meaningful features as we show below, and the smoothed separable NMF algorithms do produce better results than the non-smoothed ones.
We observe that the result is consistently improved for $p>1$ and $p$ not too large. 
SVCA outperforms ALLS in all cases.
The best solution from SVCA is always better than the solution from SSPA, and the median solution from SVCA is better than the solution from SSPA for all data sets except Umist.
In this experiment the best $p$ is the same for SVCA and SSPA in most cases.
For the data sets with fewer pixels (CBCL, Frey), a value of $p$ larger than $20$ deteriorates the performance of SVCA and SSPA. 
This makes sense since there are fewer than 600 pixels, and hence the rule of thumb that $p \ll \frac{n}{r}$ is not respected (recall that $p = \frac{n}{r}$ would be the extreme case where there are exactly $\frac{n}{r}$ data points close to each column of $W$). 
For the data sets with more pixels (ORL, UMIST), $p$ can be rather large and still leads to a decrease in the error for $p$ up to $200$. 
    
\begin{table}[ht!]
\caption{Relative reconstruction errors ($\min_{H \geq 0} \| X - WH \|_F / \| X \|_F$) in percent resulting from the processing of facial image datasets with ALLS, SVCA, and SSPA, with different values of parameter $p$. SVCA(1) and SSPA(1) are equivalent to VCA and SPA. For non-deterministic algorithms ALLS and SVCA, we show the minimum, median, standard deviation, and maximum of the error over 30 trials.}
\centering
\begin{tabular}{ll|ccc|ccc}
     &    & \multicolumn{3}{c|}{CBCL}                & \multicolumn{3}{c}{Frey}                 \\
     & p  & Min     & Med $\pm$ std        & Max     & Min     & Med $\pm$ std        & Max     \\ \hline
ALLS & 1  & 19.36 & 20.08 $\pm$ 0.33 & 20.95 & 21.05 & 22.38 $\pm$ 0.58 & 23.83 \\
     & 5  & \textbf{18.53} & 19.60 $\pm$ 0.51 & 20.41 & \textbf{20.75} & \textbf{22.32} $\pm$ 0.73 & 23.96 \\
     & 10 & 18.67 & \textbf{19.44} $\pm$ 0.78 & 21.88 & 21.30 & 22.85 $\pm$ 0.85 & 24.57 \\
     & 20 & 19.34 & 20.28 $\pm$ 1.17 & 23.83 & 22.30 & 24.05 $\pm$ 1.10 & 27.42 \\ \hline
SVCA & 1  & 19.25 & 19.81 $\pm$ 0.31 & 20.48 & 20.99 & 22.18 $\pm$ 0.64 & 24.02 \\
     & 5  & 18.72 & 19.21 $\pm$ 0.34 & 20.14 & 20.69 & 21.57 $\pm$ 0.52 & 22.85 \\
     & 10 & 18.48 & \textbf{18.84} $\pm$ 0.35 & 19.76 & \textbf{20.68} & \textbf{21.29} $\pm$ 0.43 & 22.57 \\
     & 20 & \textbf{18.44} & 18.95 $\pm$ 0.25 & 19.55 & 20.79 & 21.70 $\pm$ 0.45 & 22.71 \\ \hline
SSPA & 1  & \multicolumn{3}{c|}{20.76}             & \multicolumn{3}{c}{22.57}              \\
     & 5  & \multicolumn{3}{c|}{20.05}             & \multicolumn{3}{c}{21.65}              \\
     & 10 & \multicolumn{3}{c|}{\textbf{19.63}}             & \multicolumn{3}{c}{\textbf{21.32}}              \\
     & 20 & \multicolumn{3}{c|}{20.83}             & \multicolumn{3}{c}{21.58}             \\
     & & & & & & & \\ \hline
          &     & \multicolumn{3}{c|}{ORL}         & \multicolumn{3}{c}{Umist}        \\
     & p   & Min   & Med $\pm$ std    & Max   & Min   & Med $\pm$ std    & Max   \\ \hline
ALLS & 1   & 23.64 & 24.08 $\pm$ 0.18 & 24.35 & 15.38 & 16.00 $\pm$ 0.25 & 16.46 \\
     & 50  & \textbf{22.87} & \textbf{23.62} $\pm$ 0.42 & 24.92 & \textbf{15.34} & \textbf{15.91} $\pm$ 0.32 & 16.48 \\
     & 200 & 23.34 & 24.33 $\pm$ 0.60 & 25.93 & 16.27 & 16.96 $\pm$ 0.54 & 18.87 \\
     & 400 & 24.50 & 25.31 $\pm$ 0.45 & 26.56 & 17.28 & 18.21 $\pm$ 0.49 & 19.07 \\ \hline
SVCA & 1   & 23.34 & 24.04 $\pm$ 0.32 & 24.70 & 15.40 & 15.95 $\pm$ 0.24 & 16.28 \\
     & 50  & 22.31 & 22.89 $\pm$ 0.32 & 23.63 & 14.95 & \textbf{15.37} $\pm$ 0.28 & 16.08 \\
     & 200 & \textbf{22.22} & \textbf{22.70} $\pm$ 0.29 & 23.53 & \textbf{14.90} & \textbf{15.37} $\pm$ 0.25 & 15.91 \\
     & 400 & 22.42 & 23.00 $\pm$ 0.34 & 23.86 & 15.29 & 15.85 $\pm$ 0.30 & 16.46 \\ \hline
SSPA & 1   & \multicolumn{3}{c|}{25.76}     & \multicolumn{3}{c}{16.56}      \\
     & 50  & \multicolumn{3}{c|}{24.14}     & \multicolumn{3}{c}{15.72}      \\
     & 200 & \multicolumn{3}{c|}{\textbf{23.50}}     & \multicolumn{3}{c}{\textbf{14.98}}      \\
     & 400 & \multicolumn{3}{c|}{24.50}     & \multicolumn{3}{c}{15.40} 
\end{tabular}\label{tab:xpfaces1}
\end{table} 
Figure~\ref{fig:xpfaces2-orl} reports the relative errors for SVCA and SSPA on the data set ORL, using the mean and the median aggregation, for various values of $p$.
\begin{figure}[ht!]
    \centering

\begin{tikzpicture}
    \begin{axis}[
        width=\myfigsize\textwidth,
        scale only axis,
xlabel={Parameter $p$},
        ylabel={Relative error (\%)},
        legend style={at={(0.97,0.97)}, anchor=north east, font=\footnotesize}]
    
    \addplot[color=mygreen, mark=square] table [x index=0, y index=1] {xpfaces2-orl.dat}; \addlegendentry{SVCA-med};
    \addplot[color=mygreen, mark=triangle, dashed] table [x index=0, y index=2] {xpfaces2-orl.dat}; \addlegendentry{SVCA-mean};
    
    \addplot[color=myorange, mark=square] table [x index=0, y index=3] {xpfaces2-orl.dat}; \addlegendentry{SSPA-med};
    \addplot[color=myorange, mark=triangle, dashed] table [x index=0, y index=4] {xpfaces2-orl.dat}; \addlegendentry{SSPA-mean};
    
    \end{axis}
\end{tikzpicture}
     \caption{Results of the processing of the facial images data set ORL. Values for SVCA are the medians over 30 trials.} 
    \label{fig:xpfaces2-orl}
\end{figure}
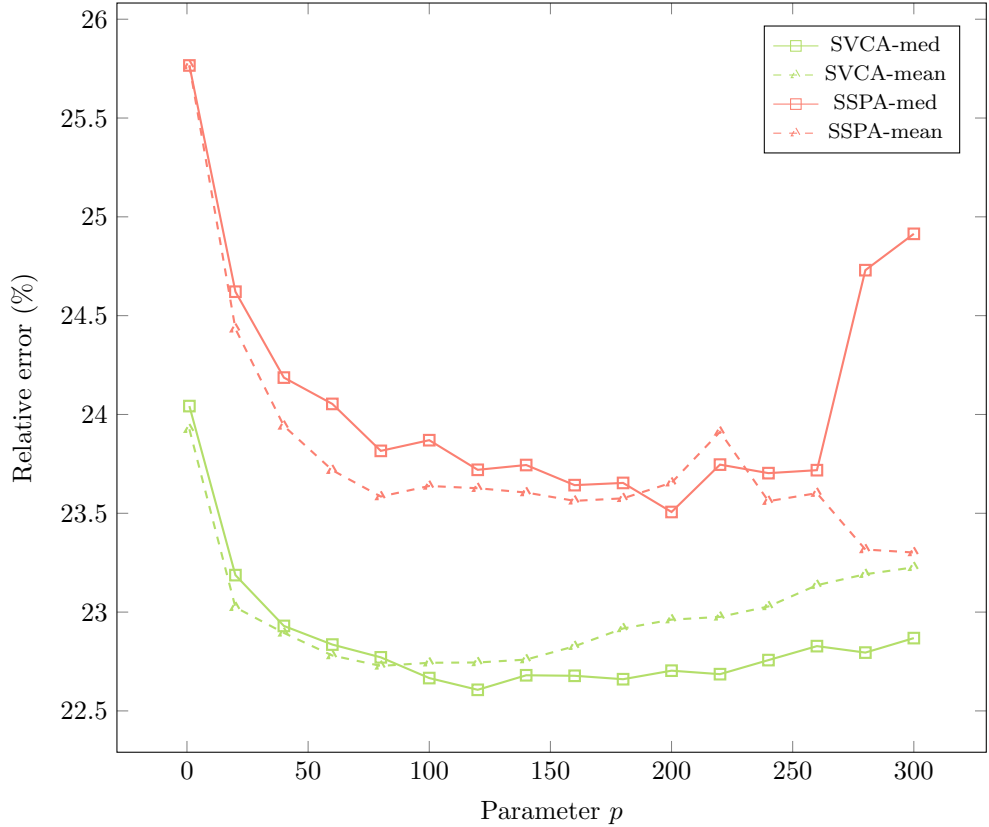
Figures for other data sets are included in the supplementary material (page~\pageref{morexp}).
We observe that, as for the synthetic data sets and hyperspectral images, the smoothed algorithms allow one to significantly and consistently reduce the error, given that $p$ is not too large.  
The difference between using the mean and the median is not significant, although the mean seems to perform slightly better for SVCA.
The reason is that there is no outlier in these data sets, and the Gaussian assumption for the noise is reasonable. 
As opposed to the experiment on hyperspectral data, here SVCA outperforms SSPA on average.

As an illustration, Figure~\ref{fig:abundancemapsorl} shows the facial features extracted by 
VCA, SVCA, SPA, SSPA.
They represent the proportion of every extracted facial feature in the pixels of the images.
We observe that SVCA and SSPA, apart from reducing the error as discussed above, also produce sharper facial features. 
These features are more clearly delimited and separated, they grasp more finely the parts of the human face such as eyes, nose, and chin. 
\begin{figure*}[ht!]
\centering
\subfloat[VCA, error$=23.35\%$]{
    \includegraphics[width=0.98\textwidth]{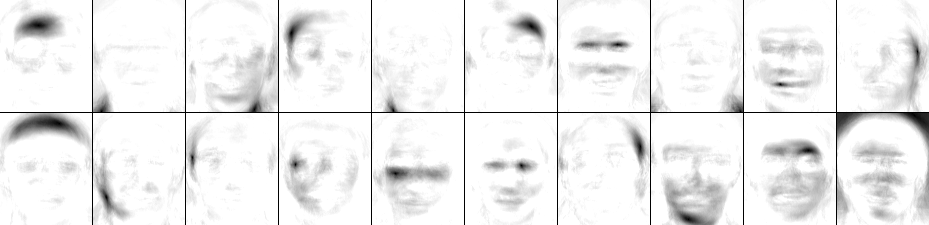}
}\\
\subfloat[SVCA $p$=120, error$=22.18\%$]{
    \includegraphics[width=0.98\textwidth]{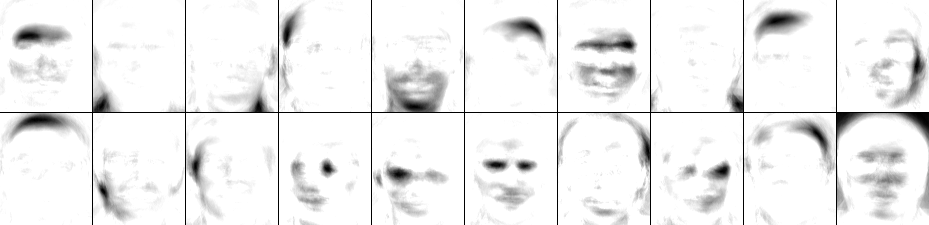}
}\\
\subfloat[SPA, error$=25.77\%$]{
    \includegraphics[width=0.98\textwidth]{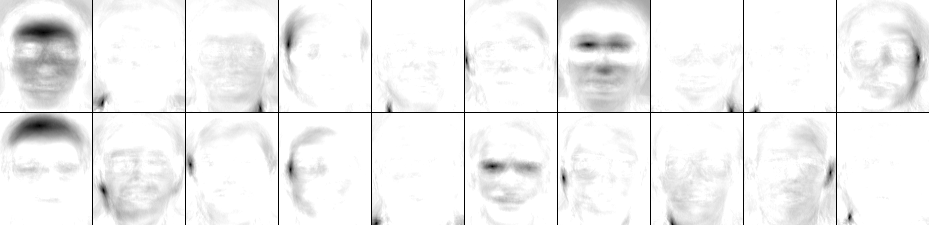}
}\\
\subfloat[SSPA $p$=200, error$=23.51\%$]{
    \includegraphics[width=0.98\textwidth]{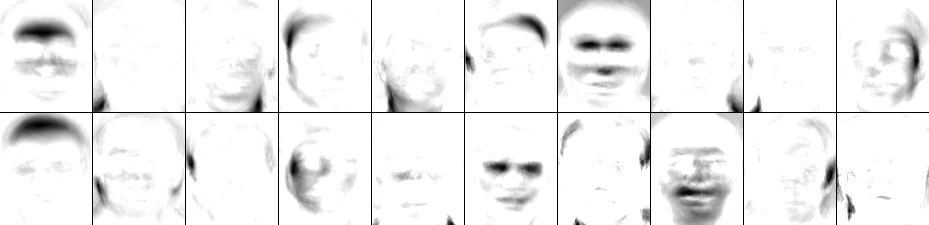}
}\\
  \caption{Abundance maps of the unmixing of the facial images data set ORL (that is, reshaped rows of $H$) with different algorithms. Features have been reordered for easier comparison. Parameters $p$ have been chosen as the best from \cref{fig:xpfaces2-orl}. Error corresponds to $\min_{H \geq 0} \| X - WH \|_F / \| X \|_F$. For VCA and SVCA, we show the best solution over 30 trials.}
  \label{fig:abundancemapsorl}
\end{figure*}

}

\newpage 

\subsection{Discussion}

Let us make some additional points regarding SVCA and SSPA. 

\paragraph*{Which algorithm should one use?} SVCA allows to generate different solutions, among which the best solution w.r.t.\ the reconstruction error can be found. Therefore, in practice and when time and resources allow, we recommend running SSPA once and SVCA several times, with different values of $p$, and keep the best solution.

\paragraph*{Which aggregation should one use?} \revise{If the noise present in the data is assumed to follow a Gaussian distribution, the mean aggregation is more adapted. 
In case of sparse noise or in the presence of outliers, then the median aggregation will perform better.}
Apart from the average and the median, other aggregation methods could perform better depending on the noise statistics and data set at hand. 
For instance, \cite{kervazo2020faster} uses an aggregation on manifold in the context of sparse matrix factorization to better take into account the structure of the columns of $W$.
\revise{Exploring more advanced aggregation methods for smoothed separable NMF would be an interesting direction for future research.}

\paragraph*{How to select $p$?} Choosing a value for the parameter $p$ is crucial and not trivial.
{Currently, one needs to have a prior knowledge on the number of data points close to the vertices, or use an empirical trial-and-error method.} 
\revise{A strategy to determine the best $p$ for a given setting is a particularly  interesting direction of research.} 
\revise{It would also be particularly meaningful to consider a different value of $p$ for every vertex}, as the number of proximal latent points typically varies for each vertex.

\paragraph*{Spectral variability in blind HU}
In blind HU, an issue with separable NMF algorithms is that they identify a single pixel to represent a material.
It is however well-known that the spectral signature of an endmember may vary across the pixels of the image, for example because of differences in light intensity or orientation. 
This is known as \emph{spectral variability}. 
By construction, most separable NMF algorithms, such as VCA and SPA, will identify pure pixels that do not represent well the average behaviour of a material, but rather a pure pixel located at the boundary of the convex hull of the variations of the spectral signature of that endmember.
Therefore, working on the smoothed data set, which averages every subset of $p$ data points, allows to better represent this average behaviour.

\paragraph*{Designing other smoothed separable NMF algorithms} 
The proximal latent points assumption could be used to generalize other separable NMF algorithms. 
In this paper, we validated the idea on the two most widely used separable NMF algorithms, namely VCA and SPA, but the same idea can be applied to any separable NMF algorithms such as SNPA~\citep{gillis2014successive}.

\section{Conclusion}

In this work, we investigated the smoothed separable NMF model, 
that strengthens the separability assumption by assuming the presence of several data points nearby each column of the basis matrix $W$. 
Inspired by the existing algorithm ALLS, we developed smoothed variants of two separable NMF algorithms, namely VCA and SPA.
Empirically, we showed that our smoothed methods outperform both the non-smoothed ones and ALLS, for both synthetic data sets and for \revise{two real-world applications, namely} the unmixing of hyperspectral images \revise{and feature extraction in facial images data sets}. 
We showed that the proximal latent points assumption is verified in hyperspectral images \revise{and in facial images data sets}, and that smoothed separable NMF algorithms are a more effective tool for \revise{these applications}.

Further works include the design of other smoothed separable NMF algorithms, and the use of SVCA and SSPA for other applications, in particular in machine learning~\citep{bhattacharyya2020finding, bakshilearning}. 
\revise{Exploring new aggregation methods and smarter ways to choose the parameter $p$ are also relevant directions for future research.}

\newpage

\section*{Acknowledgments}
\revise{We sincerely thank the reviewer for taking the time to carefully read the paper, and for the very detailed and insightful feedback that helped us improve our paper.} 

NN and NG acknowledge the support by the Fonds de la Recherche Scientifique - FNRS and the Fonds Wetenschappelijk Onderzoek - Vlanderen (FWO) under EOS Project no O005318F-RG47. NG also acknowledges the support of the Francqui Foundation.

\small 

\bibliographystyle{elsarticle-num}
\bibliography{smoothedVCA}

\newpage

\section*{Supplementary Material: Additional experiments}\label{morexp}

\newenvironment{customlegend}[1][]{%
    \begingroup
    \csname pgfplots@init@cleared@structures\endcsname
    \pgfplotsset{#1}%
}{%
    \csname pgfplots@createlegend\endcsname
    \endgroup
}%
\def\addlegendimage{\csname pgfplots@addlegendimage\endcsname}

\begin{figure*}[ht]
\centering
\pgfplotsset{myparam/.style=
  {width=0.5\textwidth,
   cycle list name=color list,
   xmin=1,
   xmax=162,
   ymax=760}}

\noindent
\begin{tikzpicture}
\matrix {
        \begin{axis}[myparam,
                    title={VCA}, 
                    ylabel={Reflectance}]
        \foreach \column in {0,...,5}{
          \addplot+[] table[x expr=\coordindex+1, y index=\column] {xp/sig/Urban_W_vca.txt};
        }
        \end{axis}
    &
        \begin{axis}[myparam,
                    title={SVCA}]
        \foreach \column in {0,...,5}{
          \addplot+[] table[x expr=\coordindex+1, y index=\column] {xp/sig/Urban_W_svca.txt};
        }
        \end{axis}
    \\
        \begin{axis}[myparam,
                    title={SPA},
                    xlabel={\# Bands},
                    ylabel={Reflectance}]
        \foreach \column in {0,...,5}{
          \addplot+[] table[x expr=\coordindex+1, y index=\column] {xp/sig/Urban_W_spa.txt};
        }
        \end{axis}
    &
        \begin{axis}[myparam,
                    title={SSPA}, 
                    xlabel={\# Bands}]
        \foreach \column in {0,...,5}{
          \addplot+[] table[x expr=\coordindex+1, y index=\column] {xp/sig/Urban_W_sspa.txt};
        }
        \end{axis}
        \\
    };
\end{tikzpicture}
\\
\begin{tikzpicture}
    \begin{customlegend}[legend entries={1,...,6}, cycle list name=color list, legend columns=6]
        \addlegendimage{color=red}
        \addlegendimage{color=blue}
        \addlegendimage{color=black}
        \addlegendimage{color=yellow}
        \addlegendimage{color=brown}
        \addlegendimage{color=teal}
    \end{customlegend}
\end{tikzpicture}
  \caption{Spectral signatures from the unmixing of the Urban hyperspectral image (that is, columns of $W$) with different algorithms. Parameters $p$ have been chosen as the best from fig.~6 of the main document. For VCA and SVCA, we show the best solution over 30 trials.
  Spectral signatures correspond to the abundance maps shown in fig.~7 of the main document, in the same order. For example, for SSPA, the spectral signatures correspond to (1)~dirt, (2)~roof tops, (3)~trees, (4)~grass, (5)~a mixture of road and dirt, and (6)~roads.}
  \label{fig:signatures}
\end{figure*}
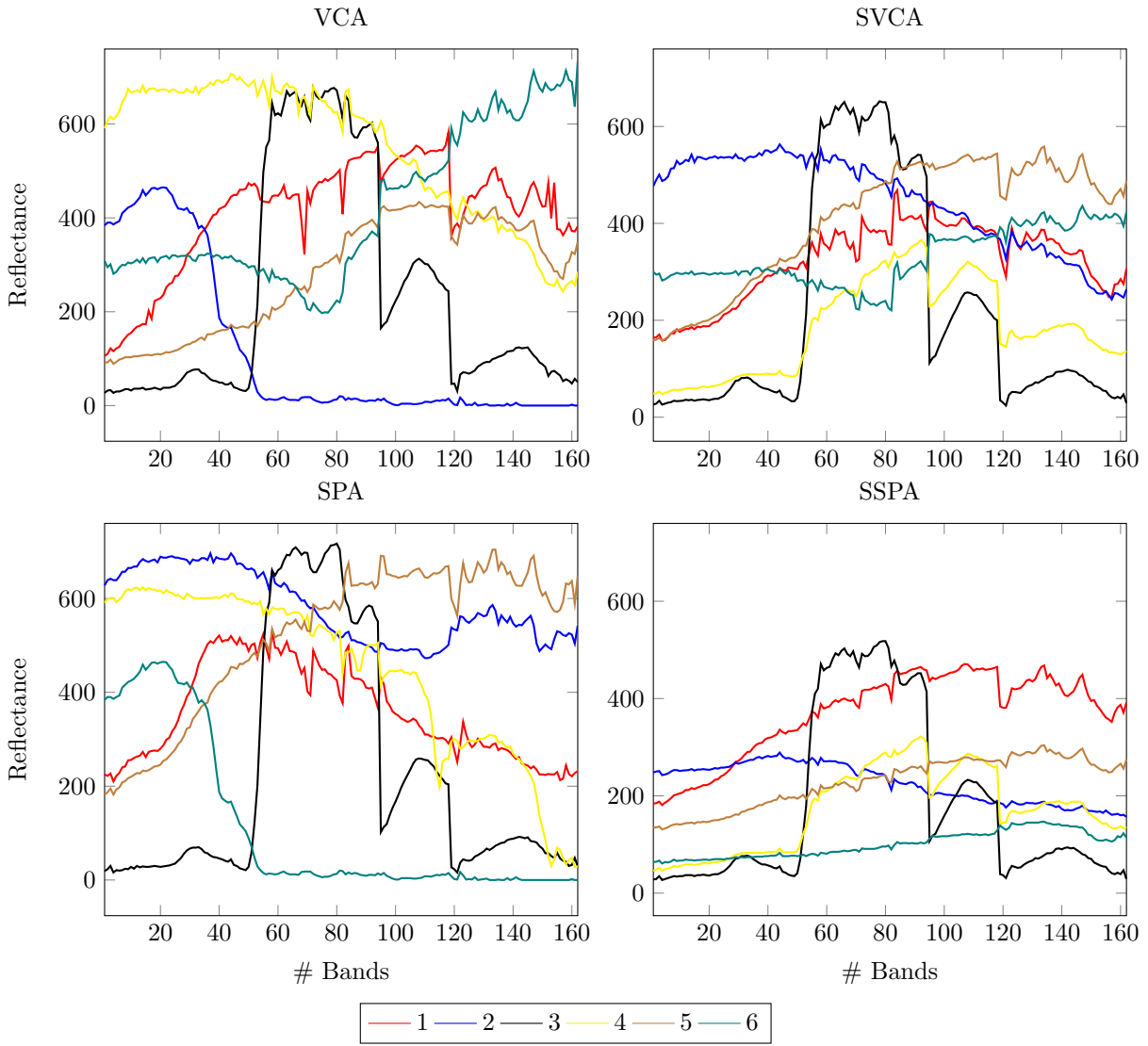

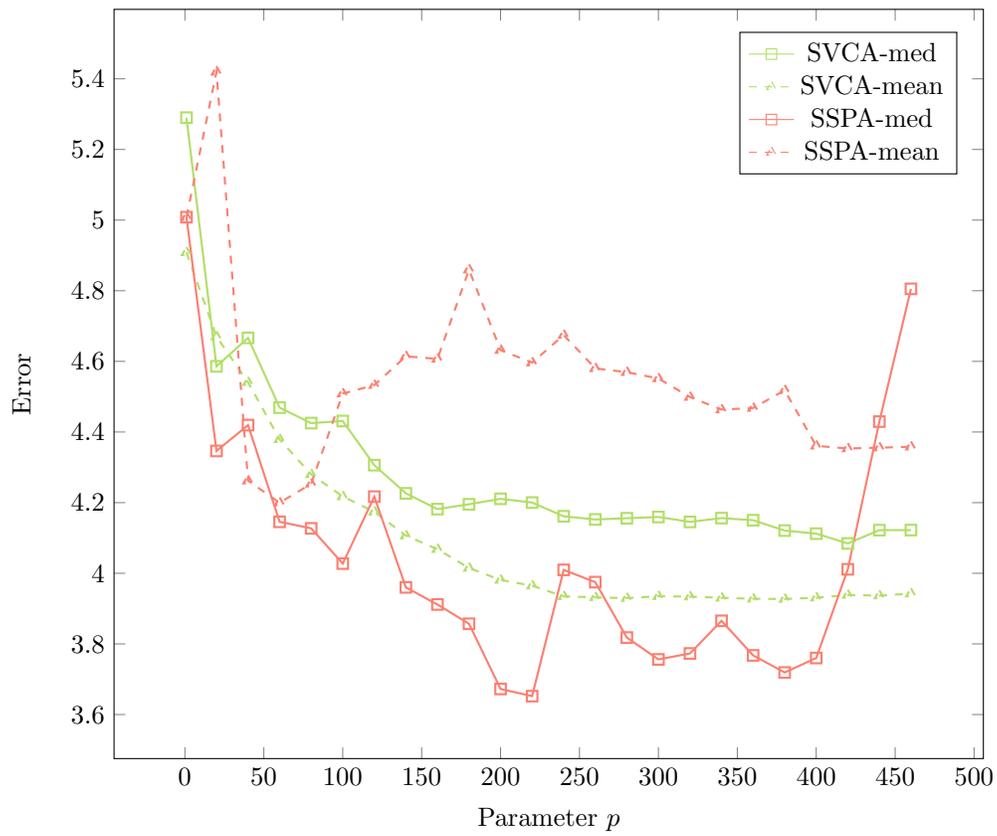
\begin{figure}[ht]
    \centering

\begin{tikzpicture}
    \begin{axis}[
        width=\myfigsize\textwidth,
        scale only axis,
        xlabel={Parameter $p$},
        ylabel={Error},
        legend style={at={(0.97,0.97)}, anchor=north east}]
    
    \addplot[color=mygreen, mark=square] table [x index=0, y index=1] {xp/xphsu2-terrainpp.dat}; \addlegendentry{SVCA-med};
    \addplot[color=mygreen, mark=triangle, dashed] table [x index=0, y index=2] {xp/xphsu2-terrainpp.dat}; \addlegendentry{SVCA-mean};
    
    \addplot[color=myorange, mark=square] table [x index=0, y index=3] {xp/xphsu2-terrainpp.dat}; \addlegendentry{SSPA-med};
    \addplot[color=myorange, mark=triangle, dashed] table [x index=0, y index=4] {xp/xphsu2-terrainpp.dat}; \addlegendentry{SSPA-mean};
    
    \end{axis}

\end{tikzpicture}
    \caption{Results of the unmixing of the hyperspectral image Terrain. Values for SVCA are the medians over 30 trials.} 
    \label{fig:xphsu2terrain}
\end{figure}

\begin{figure}[ht]
\centering
\subfloat[VCA, error$=4.03\%$]{
    \includegraphics[width=0.8\textwidth]{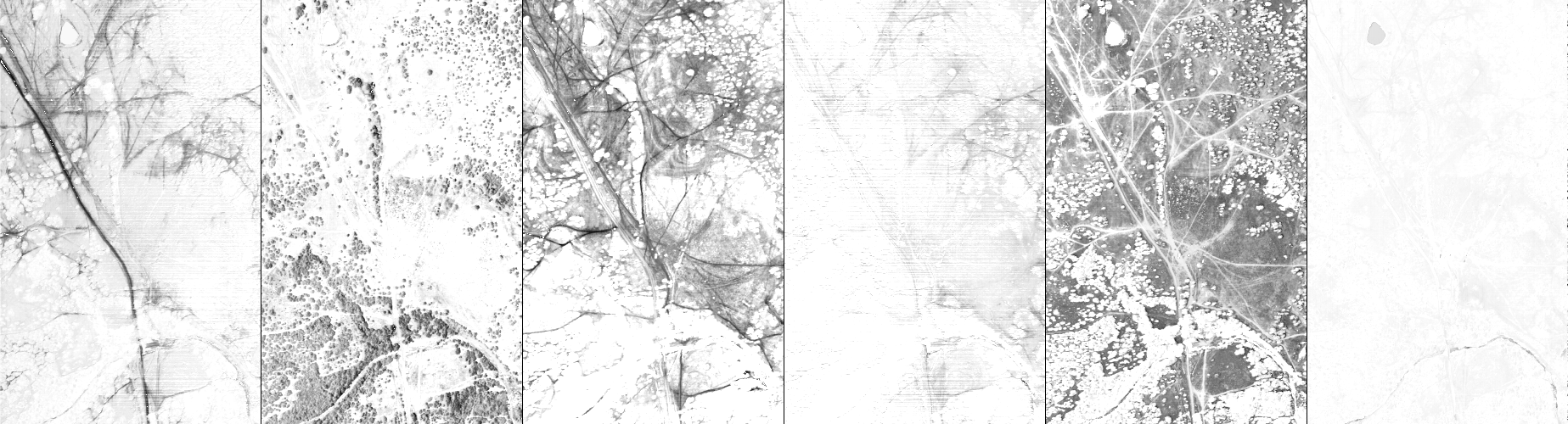}
}\\
\subfloat[SVCA $p$=420, error$=3.25\%$]{
    \includegraphics[width=0.8\textwidth]{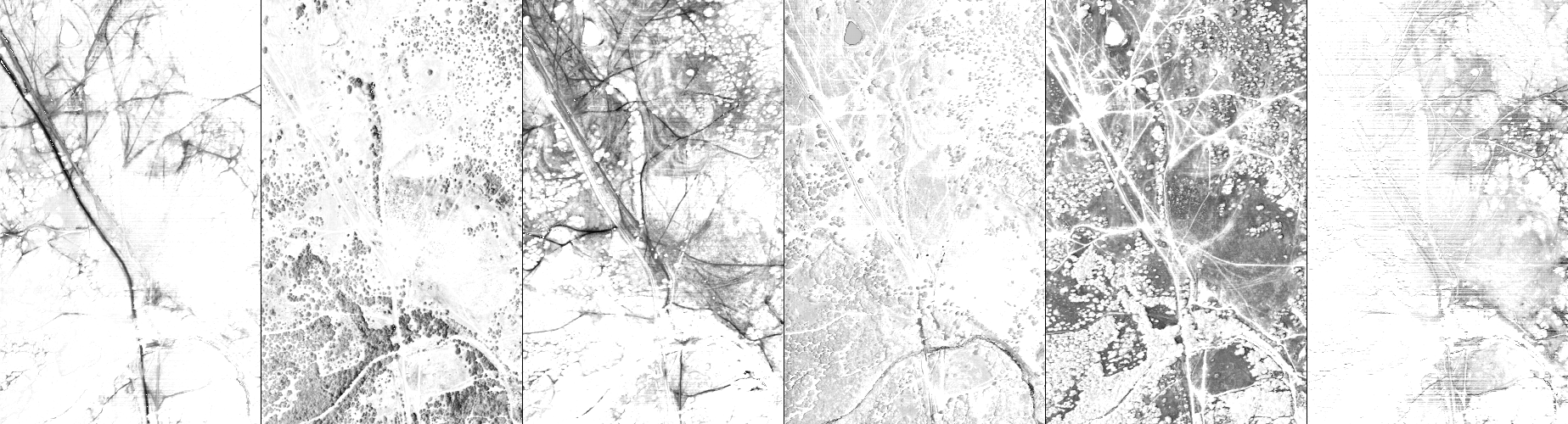}
}\\
\subfloat[SPA, error$=5.01\%$]{
    \includegraphics[width=0.8\textwidth]{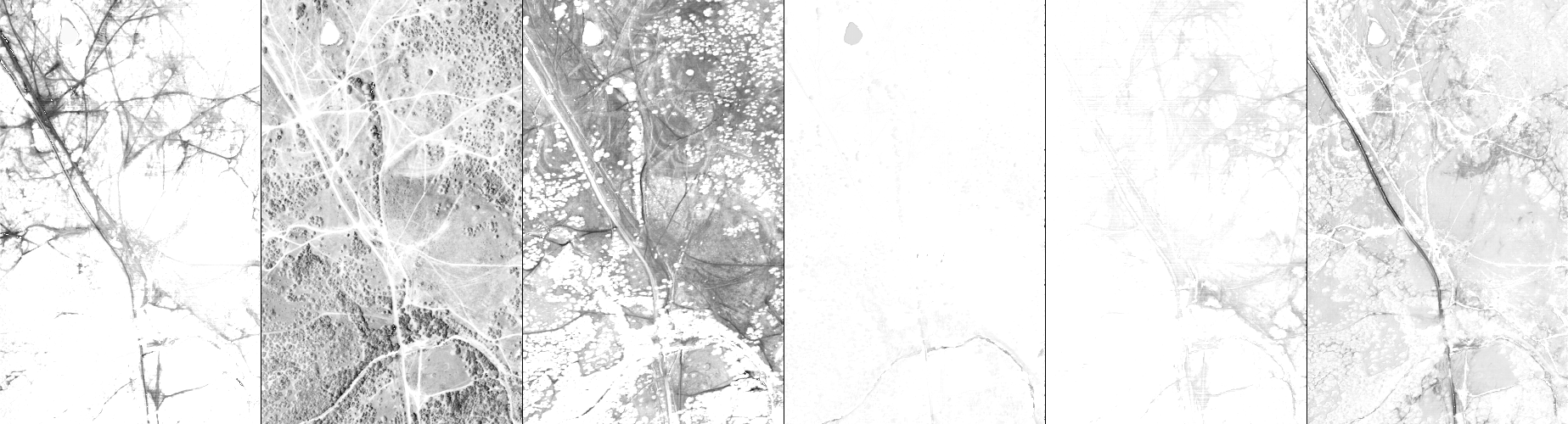}
}\\
\subfloat[SSPA $p$=220, error$=3.65\%$]{
    \includegraphics[width=0.8\textwidth]{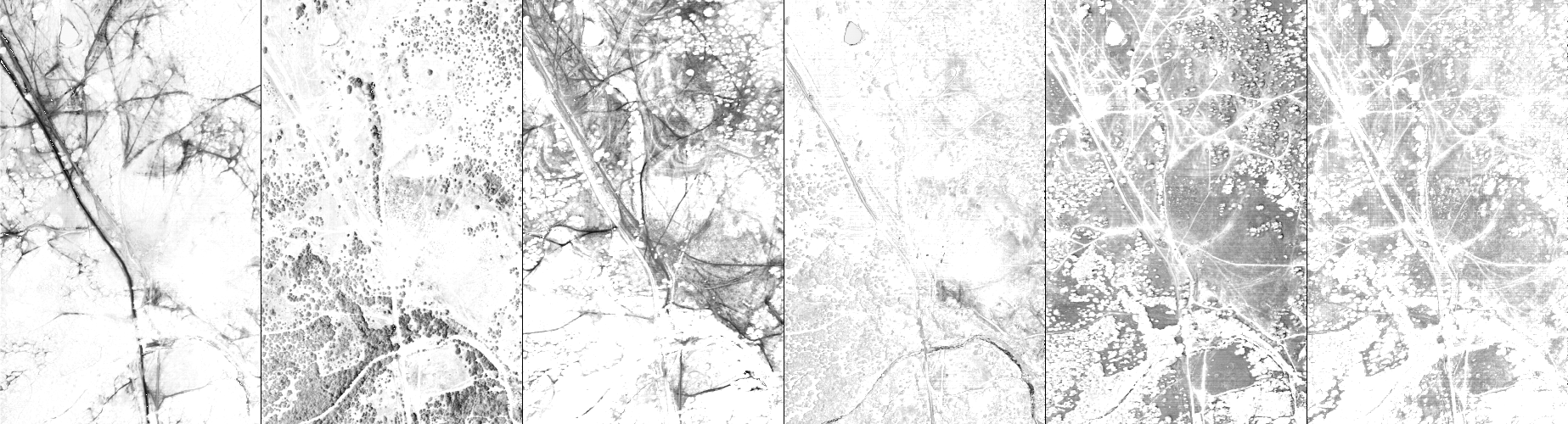}
}\\
  \caption{Abundance maps of the unmixing of the Terrain hyperspectral image (that is, reshaped rows of $H$) with different algorithms. Endmembers have been reordered for easier comparison. Parameters $p$ have been chosen as the best from \cref{fig:xphsu2terrain}. Error corresponds to $\min_{H \geq 0} \| X - WH \|_F / \| X \|_F$. For VCA and SVCA, we show the best solution over 30 trials.}
  \label{fig:abundancemapsterrain}
\end{figure}

\begin{figure}[ht]
    \centering

\begin{tikzpicture}
    \begin{axis}[
        width=\myfigsize\textwidth,
        scale only axis,
        xlabel={Parameter $p$},
        ylabel={Error},
        legend style={at={(0.97,0.97)}, anchor=north east}]
    
    \addplot[color=mygreen, mark=square] table [x index=0, y index=1] {xp/xphsu2-sandiegopp.dat}; \addlegendentry{SVCA-med};
    \addplot[color=mygreen, mark=triangle, dashed] table [x index=0, y index=2] {xp/xphsu2-sandiegopp.dat}; \addlegendentry{SVCA-mean};
    
    \addplot[color=myorange, mark=square] table [x index=0, y index=3] {xp/xphsu2-sandiegopp.dat}; \addlegendentry{SSPA-med};
    \addplot[color=myorange, mark=triangle, dashed] table [x index=0, y index=4] {xp/xphsu2-sandiegopp.dat}; \addlegendentry{SSPA-mean};
    
    \end{axis}

\end{tikzpicture}
    \caption{Results of the unmixing of the hyperspectral image San Diego. Values for SVCA are the medians over 30 trials.} 
    \label{fig:xphsu2sandiego}
\end{figure}
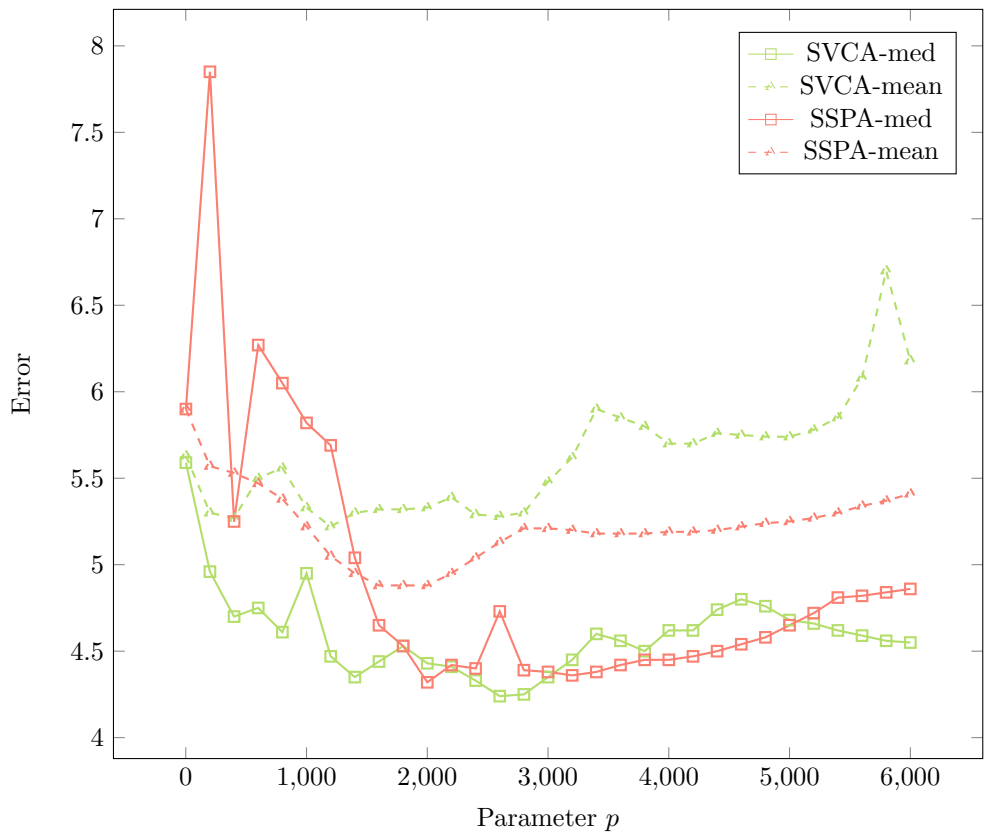

\begin{figure}[ht]
\centering
\subfloat[VCA, error$=3.95\%$]{
    \includegraphics[width=0.98\textwidth]{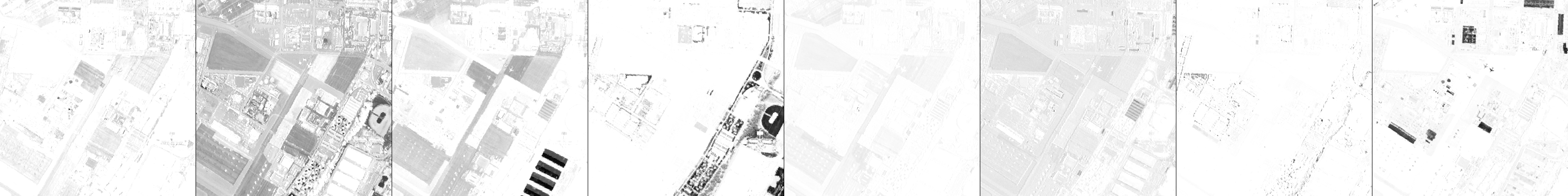}
}\\
\subfloat[SVCA $p$=2600, error$=3.74\%$]{
    \includegraphics[width=0.98\textwidth]{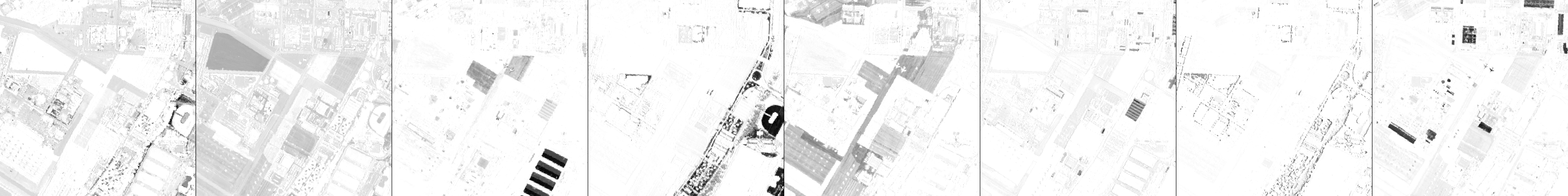}
}\\
\subfloat[SPA, error$=5.90\%$]{
    \includegraphics[width=0.98\textwidth]{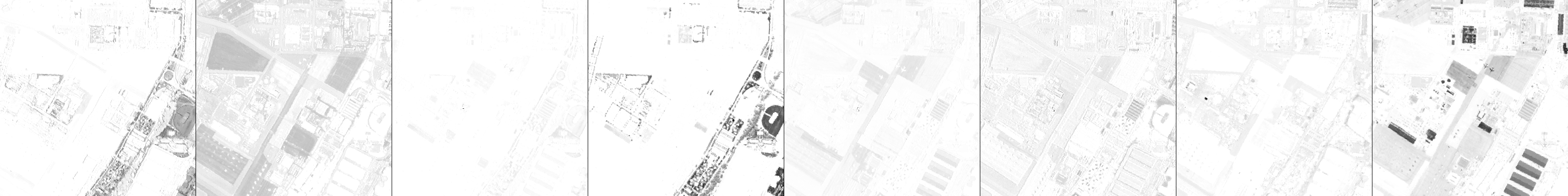}
}\\
\subfloat[SSPA $p$=2000, error$=4.32\%$]{
    \includegraphics[width=0.98\textwidth]{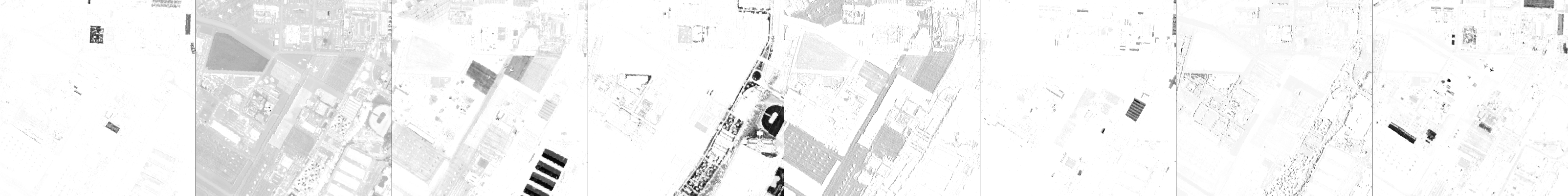}
}\\
  \caption{Abundance maps of the unmixing of the San Diego hyperspectral image (that is, reshaped rows of $H$) with different algorithms. Endmembers have been reordered for easier comparison. Parameters $p$ have been chosen as the best from \cref{fig:xphsu2sandiego}. Error corresponds to $\min_{H \geq 0} \| X - WH \|_F / \| X \|_F$. For VCA and SVCA, we show the best solution over 30 trials.}
  \label{fig:abundancemapssandiego}
\end{figure}

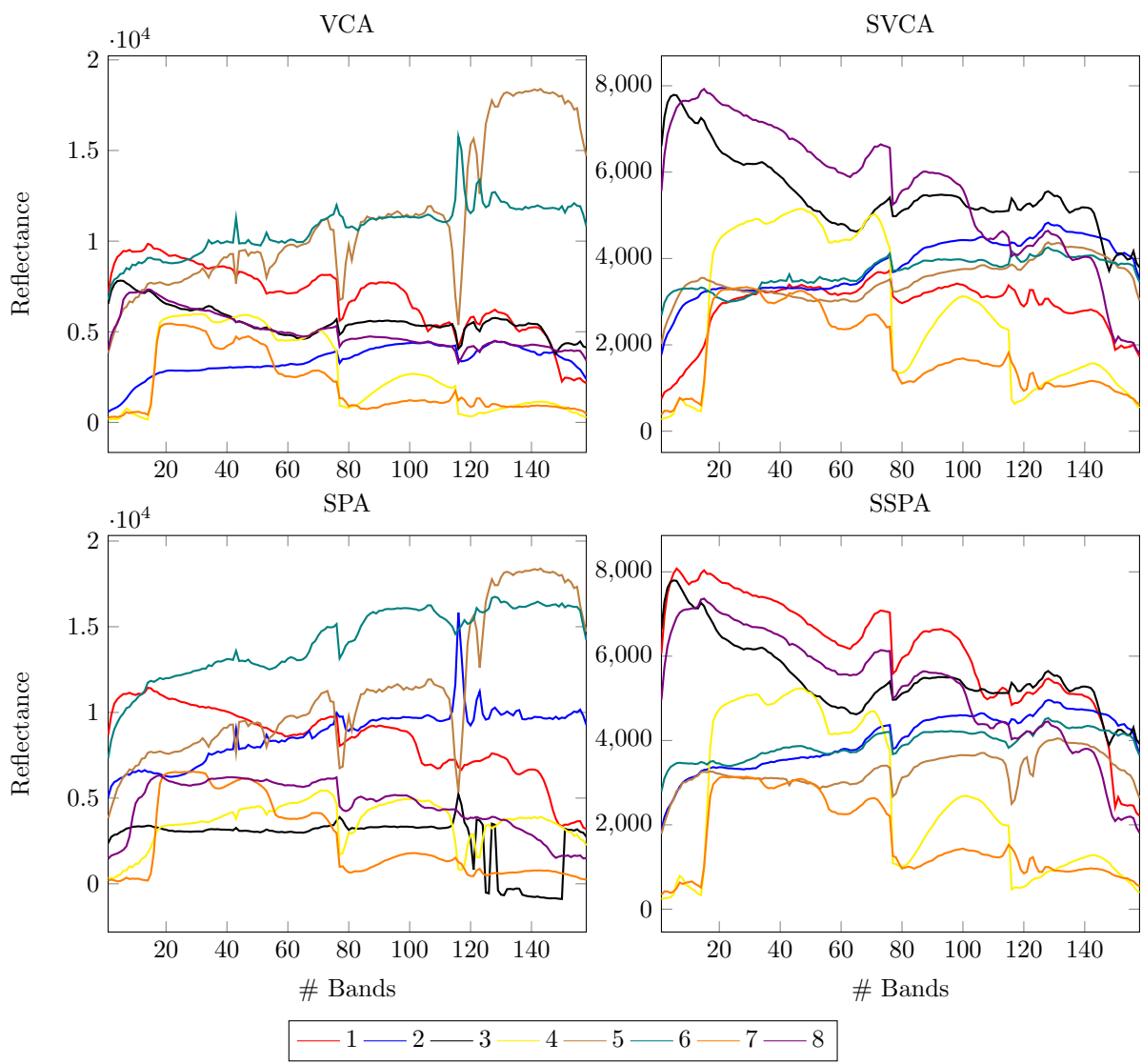
\begin{figure*}[ht]
\centering
\pgfplotsset{myparam/.style=
  {width=0.5\textwidth,
   cycle list name=color list,
   xmin=1,
   xmax=158}}

\noindent
\begin{tikzpicture}
\matrix {
        \begin{axis}[myparam,
                    title={VCA}, 
                    ylabel={Reflectance}]
        \foreach \column in {0,...,7}{
          \addplot+[] table[x expr=\coordindex+1, y index=\column] {xp/sig/SanDiego_W_vca.txt};
        }
        \end{axis}
    &
        \begin{axis}[myparam,
                    title={SVCA}]
        \foreach \column in {0,...,7}{
          \addplot+[] table[x expr=\coordindex+1, y index=\column] {xp/sig/SanDiego_W_svca.txt};
        }
        \end{axis}
    \\
        \begin{axis}[myparam,
                    title={SPA},
                    xlabel={\# Bands},
                    ylabel={Reflectance}]
        \foreach \column in {0,...,7}{
          \addplot+[] table[x expr=\coordindex+1, y index=\column] {xp/sig/SanDiego_W_spa.txt};
        }
        \end{axis}
    &
        \begin{axis}[myparam,
                    title={SSPA}, 
                    xlabel={\# Bands}]
        \foreach \column in {0,...,7}{
          \addplot+[] table[x expr=\coordindex+1, y index=\column] {xp/sig/SanDiego_W_sspa.txt};
        }
        \end{axis}
        \\
    };
\end{tikzpicture}
\\
\begin{tikzpicture}
    \begin{customlegend}[legend entries={1,...,8}, cycle list name=color list, legend columns=8]
        \addlegendimage{color=red}
        \addlegendimage{color=blue}
        \addlegendimage{color=black}
        \addlegendimage{color=yellow}
        \addlegendimage{color=brown}
        \addlegendimage{color=teal}
        \addlegendimage{color=orange}
        \addlegendimage{color=violet}
    \end{customlegend}
\end{tikzpicture}
  \caption{Spectral signatures from the unmixing of the San Diego hyperspectral image (that is, columns of $W$) with different algorithms. Parameters $p$ have been chosen as the best from \cref{fig:xphsu2sandiego}. For VCA and SVCA, we show the best solution over 30 trials.
  Spectral signatures correspond to the abundance maps shown in \cref{fig:abundancemapssandiego}, in the same order.}
  \label{fig:signaturessandiego}
\end{figure*}

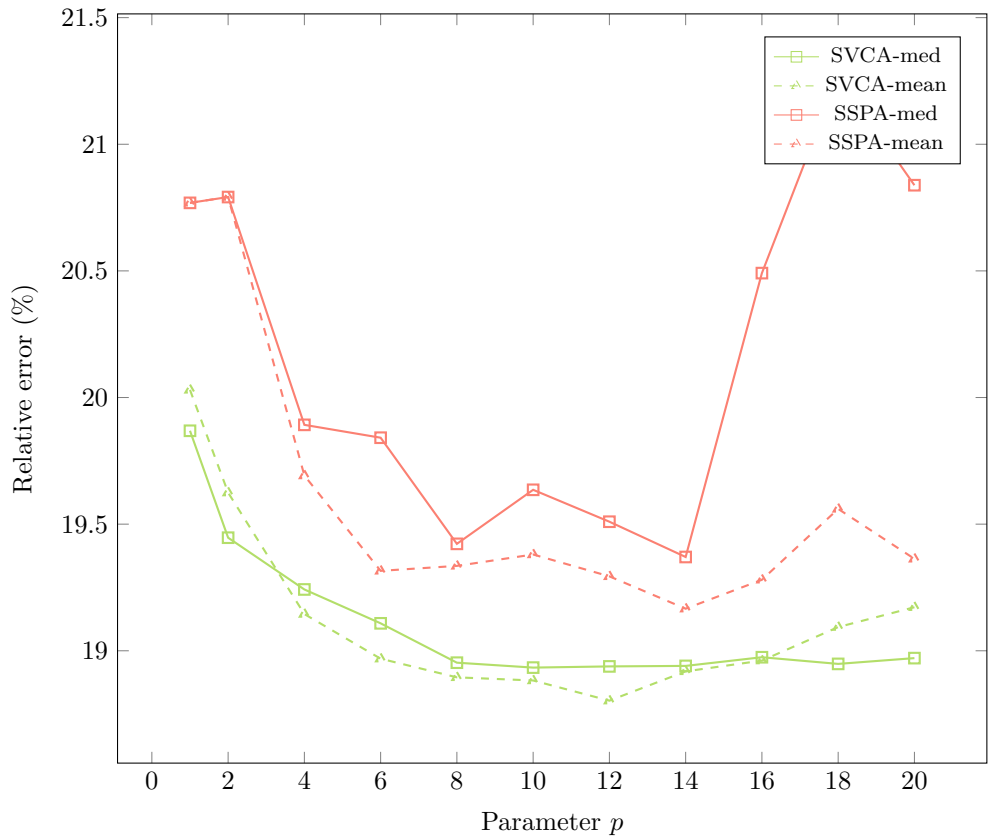
\begin{figure}[ht]
    \centering

\begin{tikzpicture}
    \begin{axis}[
        width=\myfigsize\textwidth,
        scale only axis,
        xlabel={Parameter $p$},
        ylabel={Relative error (\%)},
        legend style={at={(0.97,0.97)}, anchor=north east, font=\footnotesize}]
    
    \addplot[color=mygreen, mark=square] table [x index=0, y index=1] {xp/xpfaces2-cbcl.dat}; \addlegendentry{SVCA-med};
    \addplot[color=mygreen, mark=triangle, dashed] table [x index=0, y index=2] {xp/xpfaces2-cbcl.dat}; \addlegendentry{SVCA-mean};
    
    \addplot[color=myorange, mark=square] table [x index=0, y index=3] {xp/xpfaces2-cbcl.dat}; \addlegendentry{SSPA-med};
    \addplot[color=myorange, mark=triangle, dashed] table [x index=0, y index=4] {xp/xpfaces2-cbcl.dat}; \addlegendentry{SSPA-mean};
    
    \end{axis}
\end{tikzpicture}
    \caption{Results of the processing of the facial images dataset CBCL. Values for SVCA are the medians over 30 trials.} 
    \label{fig:xpfaces2-cbcl}
\end{figure}

\begin{figure}[ht]
    \centering

\begin{tikzpicture}
    \begin{axis}[
        width=\myfigsize\textwidth,
        scale only axis,
        xlabel={Parameter $p$},
        ylabel={Relative error (\%)},
        legend style={at={(0.97,0.97)}, anchor=north east, font=\footnotesize}]
    
    \addplot[color=mygreen, mark=square] table [x index=0, y index=1] {xp/xpfaces2-frey.dat}; \addlegendentry{SVCA-med};
    \addplot[color=mygreen, mark=triangle, dashed] table [x index=0, y index=2] {xp/xpfaces2-frey.dat}; \addlegendentry{SVCA-mean};
    
    \addplot[color=myorange, mark=square] table [x index=0, y index=3] {xp/xpfaces2-frey.dat}; \addlegendentry{SSPA-med};
    \addplot[color=myorange, mark=triangle, dashed] table [x index=0, y index=4] {xp/xpfaces2-frey.dat}; \addlegendentry{SSPA-mean};
    
    \end{axis}
\end{tikzpicture}
    \caption{Results of the processing of the facial images dataset Frey. Values for SVCA are the medians over 30 trials.} 
    \label{fig:xpfaces2-frey}
\end{figure}
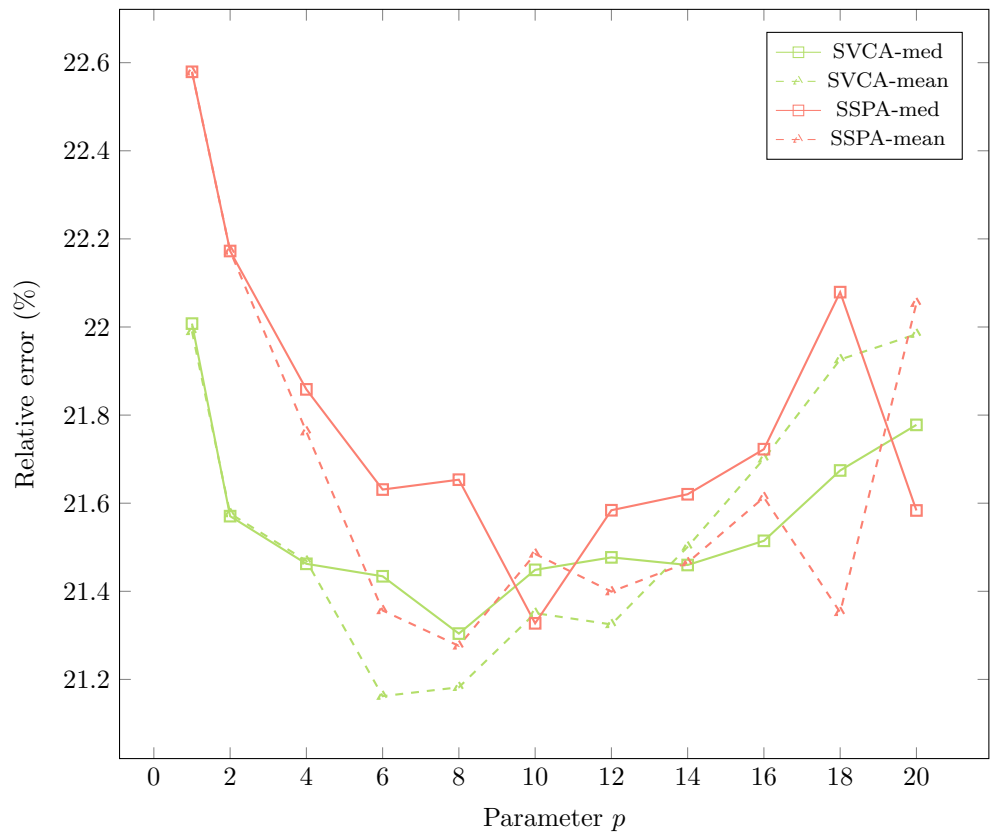

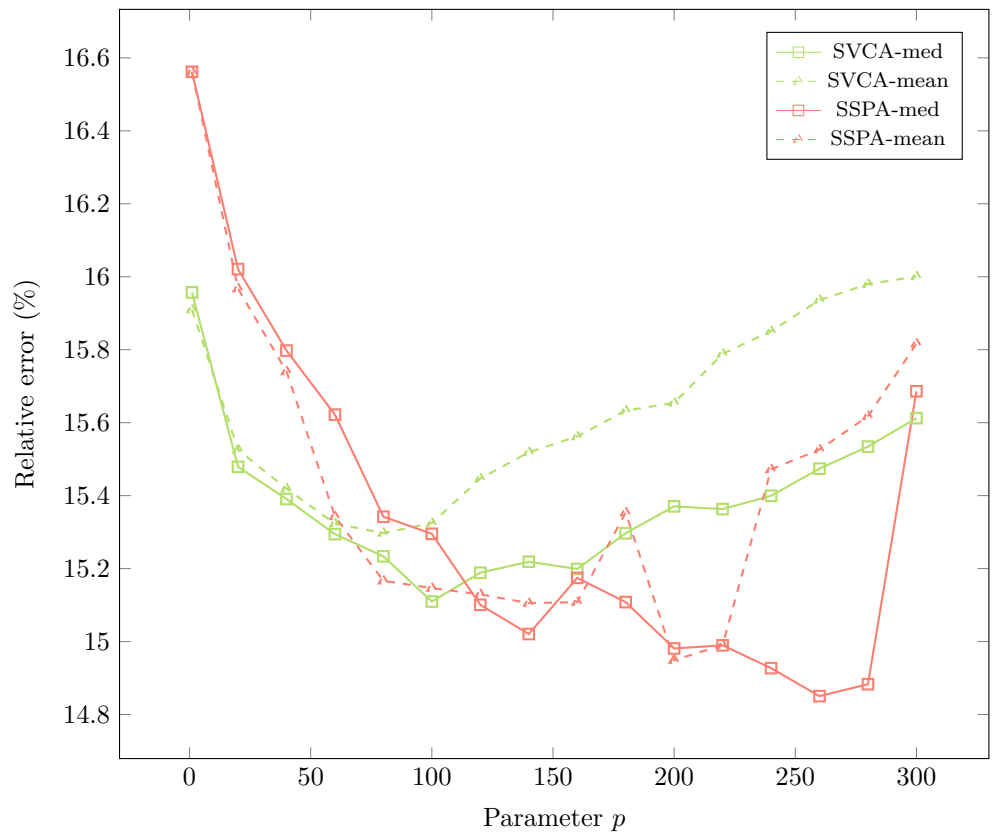
\begin{figure}[ht]
    \centering

\begin{tikzpicture}
    \begin{axis}[
        width=\myfigsize\textwidth,
        scale only axis,
        xlabel={Parameter $p$},
        ylabel={Relative error (\%)},
        legend style={at={(0.97,0.97)}, anchor=north east, font=\footnotesize}]
    
    \addplot[color=mygreen, mark=square] table [x index=0, y index=1] {xp/xpfaces2-umist.dat}; \addlegendentry{SVCA-med};
    \addplot[color=mygreen, mark=triangle, dashed] table [x index=0, y index=2] {xp/xpfaces2-umist.dat}; \addlegendentry{SVCA-mean};
    
    \addplot[color=myorange, mark=square] table [x index=0, y index=3] {xp/xpfaces2-umist.dat}; \addlegendentry{SSPA-med};
    \addplot[color=myorange, mark=triangle, dashed] table [x index=0, y index=4] {xp/xpfaces2-umist.dat}; \addlegendentry{SSPA-mean};
    
    \end{axis}
\end{tikzpicture}
    \caption{Results of the processing of the facial images dataset UMIST. Values for SVCA are the medians over 30 trials.} 
    \label{fig:xpfaces2-umist}
\end{figure}

In this appendix, we provide the spectral signatures from our experiment on the Urban hyperspectral image  (\cref{fig:signatures}).
We also provide the results from our experiments on the following hyperspectral images: Terrain (\cref{fig:xphsu2terrain,fig:abundancemapsterrain,fig:signaturesterrain}) and San Diego (\cref{fig:xphsu2sandiego,fig:abundancemapssandiego,fig:signaturessandiego}).
We also provide the results from our experiments on the following facial images datasets: CBCL (\cref{fig:xpfaces2-cbcl}), Frey (\cref{fig:xpfaces2-frey}), and UMIST (\cref{fig:xpfaces2-umist}).

\begin{figure*}[ht]
    \centering
    \pgfplotsset{myparam/.style=
  {width=0.5\textwidth,
   cycle list name=color list,
   xmin=1,
   xmax=162}}

\noindent
\begin{tikzpicture}
\matrix {
        \begin{axis}[myparam,
                    title={VCA}, 
                    ylabel={Reflectance}]
        \foreach \column in {0,...,5}{
          \addplot+[] table[x expr=\coordindex+1, y index=\column] {xp/sig/Terrain_W_vca.txt};
        }
        \end{axis}
    &
        \begin{axis}[myparam,
                    title={SVCA}]
        \foreach \column in {0,...,5}{
          \addplot+[] table[x expr=\coordindex+1, y index=\column] {xp/sig/Terrain_W_svca.txt};
        }
        \end{axis}
    \\
        \begin{axis}[myparam,
                    title={SPA},
                    xlabel={\# Bands},
                    ylabel={Reflectance}]
        \foreach \column in {0,...,5}{
          \addplot+[] table[x expr=\coordindex+1, y index=\column] {xp/sig/Terrain_W_spa.txt};
        }
        \end{axis}
    &
        \begin{axis}[myparam,
                    title={SSPA}, 
                    xlabel={\# Bands}]
        \foreach \column in {0,...,5}{
          \addplot+[] table[x expr=\coordindex+1, y index=\column] {xp/sig/Terrain_W_sspa.txt};
        }
        \end{axis}
        \\
    };
\end{tikzpicture}
\\
\begin{tikzpicture}
    \begin{customlegend}[legend entries={1,...,6}, cycle list name=color list, legend columns=6]
        \addlegendimage{color=red}
        \addlegendimage{color=blue}
        \addlegendimage{color=black}
        \addlegendimage{color=yellow}
        \addlegendimage{color=brown}
        \addlegendimage{color=teal}
    \end{customlegend}
\end{tikzpicture}
  \caption{Spectral signatures from the unmixing of the Terrain hyperspectral image (that is, columns of $W$) with different algorithms. Parameters $p$ have been chosen as the best from \cref{fig:xphsu2terrain}. For VCA and SVCA, we show the best solution over 30 trials.
  Spectral signatures correspond to the abundance maps shown in \cref{fig:abundancemapsterrain}, in the same order.}
  \label{fig:signaturesterrain}
\end{figure*}
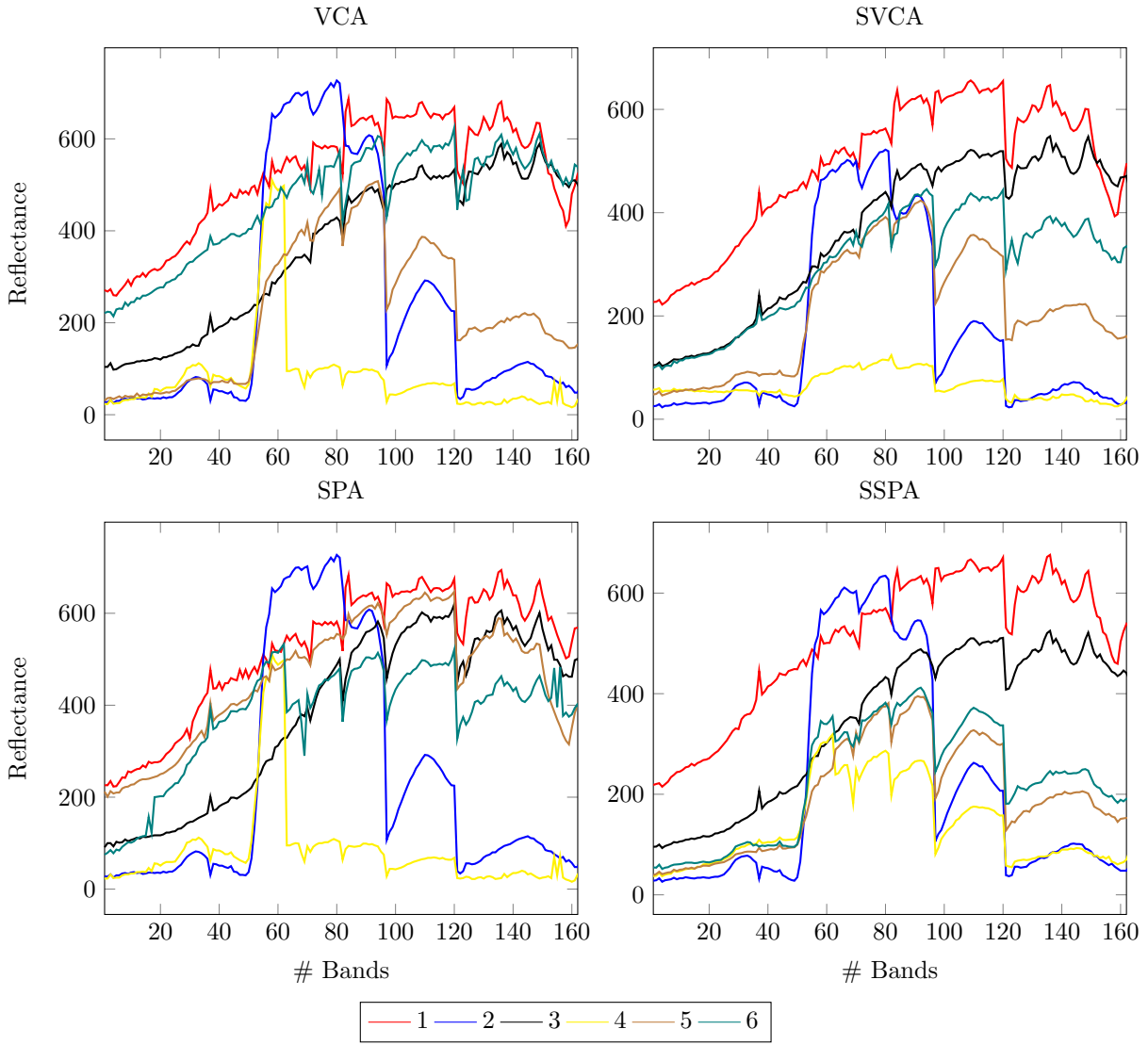


\end{document}